\documentclass[]{aastex631}
\usepackage{amsmath,amssymb}
\usepackage{graphicx}
\usepackage{graphics}
\usepackage{dcolumn}
\usepackage{bm}
\usepackage{blindtext}
\usepackage{mathptmx}
\usepackage{etoolbox}
\usepackage{color}
\usepackage{epstopdf} 
\usepackage{caption}
\usepackage{subcaption}
\definecolor{bluepig}{rgb}{0.2, 0.2, 0.6}
\definecolor{bluencs}{rgb}{0.0, 0.53, 0.74}
\definecolor{darkcerulean}{rgb}{0.03, 0.27, 0.49}
\definecolor{brightpink}{rgb}{1.0, 0.0, 0.5}
\newcommand{\markup}[1]{\textcolor{black}{#1}}

\newcommand{\Losseqn}{\mathcal{L}_{eqn}}
\newcommand{\LossBC}{\mathcal{L}_{BC}}
\newcommand{\Lossini}{\mathcal{L}_{ini}}
\newcommand{\Lossdata}{\mathcal{L}_{data}}

\usepackage{lipsum}
\usepackage{natbib}

\setcitestyle{authoryear,open={(},close={)}}

\graphicspath{{./}}
\begin{document}

\title{Physics-informed data based neural networks for two-dimensional turbulence}

\author{Vijay Kag}
\affiliation{Department of Physics, Indian Institute of Technology, Kharagpur, India}
\author{Kannabiran Seshasayanan}%
\affiliation{Department of Physics, Indian Institute of Technology, Kharagpur, India}

\author{Venkatesh Gopinath}
 \altaffiliation{Corresponding author \\
 \textit{email address}: gopinath.venkatesh2@in.bosch.com \\ \\ \\ \\
 \noindent\textit{{Preprint submitted to Physics of Fluids}}}
\affiliation{Bosch Research and Technology Center - India,  Bangalore, India
}%

\begin{abstract}
Turbulence remains a problem that is yet to be fully understood, with experimental and numerical studies aiming to fully characterise the statistical properties of turbulent flows. Such studies require huge amount of resources to capture, simulate, store and analyse the data. In this work, we present physics-informed neural network (PINN) based methods to predict flow quantities and features of two-dimensional turbulence with the help of sparse data in a rectangular domain with periodic boundaries. While the PINN model can reproduce all the statistics at large scales, the small scale properties are not captured properly. 
We introduce a new PINN model that can effectively capture the energy distribution at small scales performing better than the standard PINN based approach. It relies on the training of the low and high wavenumber behaviour separately leading to a better estimate {\markup{of the full turbulent flow}}. With $0.1 \%$ training data, we observe that the new PINN model captures the turbulent field at inertial scales leading to a general agreement of the kinetic energy spectra upto eight to nine decades as compared with the solutions from direct numerical simulation (DNS). We further apply these techniques to successfully capture the statistical behaviour of large scale modes in the turbulent flow. We believe such methods to have significant applications in enhancing the retrieval of existing turbulent data sets at even shorter time intervals. 

\end{abstract}


\section{Introduction}

Turbulent flows occur widely in many different fields, from industrial flows to naturally occurring flows such as in geophysical and astrophysical systems where energy is distributed across a continuous range of scales. Understanding and controlling turbulence is also of paramount importance in many industrial applications such as in aerospace, automotive domains, to name a few. DNS has been a mainstay for simulating turbulent flows in such systems. We generally rely upon generating large amount of data before we can understand qualitative and quantitative features of turbulence. Nevertheless, limitation of memory and computing power persists  at large Reynolds numbers and {\markup{studying}} the statistical properties of turbulence requires long duration of simulations. Experimental studies can achieve much higher Reynolds numbers and can study turbulence for longer duration, but limitations exist on acquiring velocity and pressure fields over the whole domain. 

Recently there has been a surge of data based approaches that are used for studying turbulence \citep{kim20,kochkov21,Taghi21,ghanesh21} that is gaining importance as an alternative or as an enhancement to the existing methods. A recent work by \cite{duraisamy19} reviews many methods based on machine learning for turbulence modelling and remarked that there will be a surge in data-driven modelling. Standard computational fluid dynamic (CFD) models, especially direct numerical simulations (DNS) demand high computational cost as they involve resolving turbulence down to its smallest scales. 

Usually data based approaches do not take into account the underlying physical laws of the system. Many a times, these data-based approaches also suffer drawbacks of sparse training data. Even techniques like Gaussian Process Regression (GPR) can face challenges when the system is highly nonlinear as observed in \cite{raissi17}. One such method which alleviates these drawbacks and limitations is the physics-informed neural network (PINN) which was introduced by \cite{raissi17_1,raissi17_2,raissi19,nature21}. The information of the governing equations of the system along with the boundary and initial conditions are used in the definition of the loss function which the neural network tries to minimise. The effect of adding such additional information leads to the penalizing of nonphysical solutions. It has been successfully shown to perform well in recent studies \citep{bloodflow21,nsfnets21,wang21}. Another concept of using separate PINN for each set of governing equations is also shown to perform better than a single PINN model for the entire system \citep{laubscher21}.  While PINN models have successfully reproduced steady or time periodic behaviour of different systems, their performance in the chaotic or turbulent regime is still not fully explored. In fact, \cite{nature21} mentions the limitation of PINN in studying multiscale phenomena and they suggest that new techniques are required such as xPINNs \citep{ameya20,xpinns} to handle such problems. 

Standard PINN based methods for turbulence therefore has shown difficulty in reproducing DNS solutions, especially when the {\markup{Reynolds number is high}}. The main difficulty arises from the necessity of large training points and neural network architectures which demand models that can run only on multi-GPU resources \citep{nsfnets21}. 
One possible way to overcome this difficulty is to use some training data points from existing solution from experiments or CFD simulations \citep{bloodflow21}. Recently, turbulent flows \citep{nsfnets21,lucor22} in reduced domains have been studied using PINN based methods with a small percentage of training data. In the work of \cite{nsfnets21}, such a PINN model is used for solving a sub-domain of a fully turbulent channel flow problem. They clearly show that PINN can be used in such cases to sustain turbulence over long periods of time and can also reproduce some of the turbulent features. \cite{cai2021} also use PINN to recover time evolving flow fields based on experimental data. Alternatively, PINN has also been used to reduce the amount of training data \citep{Gao21,Liu20,fukami21} to produce super-resolution solutions of low-resolution turbulence simulation data using deep learning methods.

The question still remains whether PINN based methods can capture the global behaviour of turbulence. {\markup{More specifically, whether such methods can resolve and capture the fluctuations across many spatial and temporal scales in a turbulent flow is unknown}}. Along with existing standard CFD approaches, these methods can augment them by improving the computational efficiency. With these expectations in mind, we use the methodology of PINN for studying and understanding two-dimensional turbulence. We characterize the turbulent solutions {\markup{obtained through the PINN}} based methods and show how such methods can be used to study large scale temporal dynamics. We are interested in developing a neural network that can reproduce the flow features by providing sparse training data which {\markup{comprises data from initial}} conditions, boundary conditions and some spatio-temporal points from the interior domain of the DNS solution, similar to the method used in \cite{bloodflow21}, \cite{lucor22}. Apart from this approach which we call PINN-1, we also introduce another way of training the neural network using spectral decomposition of training data which we name as PINN-2. We then proceed to demonstrate solutions of these PINN based methods which can satisfactorily reproduce the statistics of two-dimensional turbulent flows. We will also discuss where the results differ from DNS solutions and how such PINN based methods can be useful to study turbulent flows. 




\section{Methods}

\subsection{\label{sec:level1}Physics-informed neural networks} \label{Sec:Methods_PINN}

With the advent of PINN \citep{raissi19}, the physical laws of the fluid dynamical system, namely the Navier-Stokes equations are enforced as constraints on the neural network. This has proven to be a powerful technique to obtain reasonably accurate solutions when compared to traditional CFD solvers \citep{raissi19,hfm20,nsfnets21}. They particularly do well even when the problem setup has missing boundary conditions or parameters using training data points, from experiments or simulations \citep{bloodflow21,lucor22}. In this work, we use the PINN framework to study turbulence in a two-dimensional domain. The boundary conditions are kept periodic in this study but it can be easily extended to other types of domains and boundary conditions. Apart from its physical significance to the problem at hand, we were also motivated to look at such a boundary condition for turbulent flow in the PINN framework based on the remarks in \cite{nsfnets21}.

The general Navier-Stokes equation describing flow physics has the following structure \citep{batchelor_2000}, written in terms of the velocity variable ${\bf u}$ in a domain $\Omega$ as,
\begin{align}
    \partial_t {\bf u} = \mathcal{M} {\bf u} + \mathcal{N} ( {\bf u} ),
    \label{eqn:Gen_NS}
\end{align}
where $\mathcal{N}$ is the nonlinear operator and $\mathcal{M}$ is the linear operator. {\markup{Here we consider the Navier-Stokes equations for the incompressible fluid, where, {\markup{the pressure field satisfies}} a Poisson equation. The pressure term is taken to a part of $\mathcal{N} ( {\bf u} )$}. ${\bf u} ({\bf x}, 0)$ is the initial condition and the equations satisfy a set of boundary conditions denoted by $BC$. Using the technology of automatic differentiation via the neural network graph structure \citep{optbook}, the linear and nonlinear terms are calculated. For accurate neural network based approximation of the solution, a loss function has to be minimized \citep{raissi19,nsfnets21}. Although we have a well-posed problem, we take some sparse training data from the interior of the domain $\Omega$ as we are trying to model flows at high Reynolds number. Such flows will involve fluctuations across different spatial and temporal scales leading to a necessity to train the neural network with large initial and boundary training data. This leads to the requirement of more computational resources, for example, codes that can run on multi-GPUs \citep{nsfnets21}. Hence, we try to overcome this difficulty by providing training data from the interior of the problem domain $\Omega$ \citep{lucor22}. 

The equation loss $\Losseqn$ is given as: 
\begin{equation}
    \Losseqn = \frac{1}{N_{eqn}} \sum_{n=1}^{N_{eqn}} |\partial_t {\bf u} ({\bf x}_n^e, t_n^e) - \mathcal{M} {\bf u} ({\bf x}_n^e, t_n^e) - \mathcal{N}({\bf u} ({\bf x}_n^e, t_n^e))|^2,
\end{equation}
where $({\bf x}^e, t^e)$ denote the collocation points where equation loss is calculated and $N_{eqn}$ denotes the total number of these collocation points. The boundary and initial condition loss functions are given as: 
\begin{align}
    \LossBC  = & \frac{1}{N_{BC}} \sum_{n = 1}^{N_{BC}} G ( {\bf u} ({\bf x}^b_n, t^b_n) ), \\
    \Lossini = & \frac{1}{N_{ini}} \sum_{n = 1}^{N_{ini}} |{\bf u} ( {\bf x}^i_n, 0 ) - {\bf u}^e ({\bf x}^i_n, 0)|^2,  \label{eqn:loss_ini}
\end{align}
where $({\bf x}^b, t^b)$ are collocation points on the boundaries where the boundary loss is trained and $({\bf x}^i, t^i)$ are collocation points where the initial condition loss is trained. 
$N_{BC}$ denotes the number of collocation points on the boundaries and $N_{ini}$ denotes the number of collocation points at initial time. ${\bf u}^e ({\bf x}, 0)$ is the initial conditions for the set of equations in \eqref{eqn:Gen_NS}. 
{\markup{The function $G$ is a measure of the deviation of the predicted solution from the imposed boundary conditions.  It takes only positive real values for any velocity field}}.
For the current problem of two-dimensional turbulence in a periodic box, the function $G$ can be defined as follows,
\begin{align}
    & G( {\bf u} ({\bf x}^b_n, t^b_n) ) = |{\bf u} ({\bf x}^b_n, t^b_n) - {\bf u} ({\bf x}^b_n + L \hat{\bf e},t^b_n)  |^2 +  \nonumber \\
    & | {\bm \nabla}  {\bf u} ({\bf x}^b_n, t^b_n) - {\bm \nabla} {\bf u} ({\bf x}^b_n + L \hat{\bf e}, t^b_n) |^2, 
\end{align}
where {\markup{$L$ represents the domain length along both $x, y$ directions}} and $\hat{\bf e}$ represents the unit vector which is along the $x$ or the $y$ direction depending on the boundary point ${\bf x}_n^b$. 
The final loss that we compute is the loss data $\Lossdata$ which measures the deviation of the predicted solution from the exact solution at the interior points in the domain. 
It is defined as,
\begin{align}
\Lossdata = \frac{1}{N_{data}} \sum_{n=1}^{N_{data}}  |{\bf u} ({\bf x}^d_n, t^d_n) - {\bf u}^{e} ({\bf x}^d_n, t^d_n)|^2, \label{eqn:loss_data}
\end{align}
where $({\bf x}^d, t^d)$ are collocation points for training the system on a sparse set of interior points and ${\bf u}^{e}$ denotes the exact solution which is obtained either from DNS or from experimental data. 
$N_{data}$ is the number of collocation points for training on data of interior points. The total loss is thus given by:
\begin{align}
    \mathcal{L}_{tot} = \lambda_1 \Lossini + \lambda_2 \Lossdata + \lambda_3 {\markup{\LossBC + \lambda_4 \Losseqn,}}
    \label{eqn:Loss_def}
\end{align}
{\markup{where}} $\lambda_1$, $\lambda_2$, $\lambda_3$ and $\lambda_4$ are weight coefficients for the contribution from initial, data, {\markup{boundary and equation losses}} respectively. A schematic is shown in Fig. \ref{fig:schema}, where we show the $(x,y,t)$ domain of the flow field where the solution is computed using the PINN model. The collocation points where the different losses are computed are shown in different symbols and colors. 

\begin{figure}
\centering
\includegraphics[width=0.5\textwidth,height=0.5\textheight,keepaspectratio]{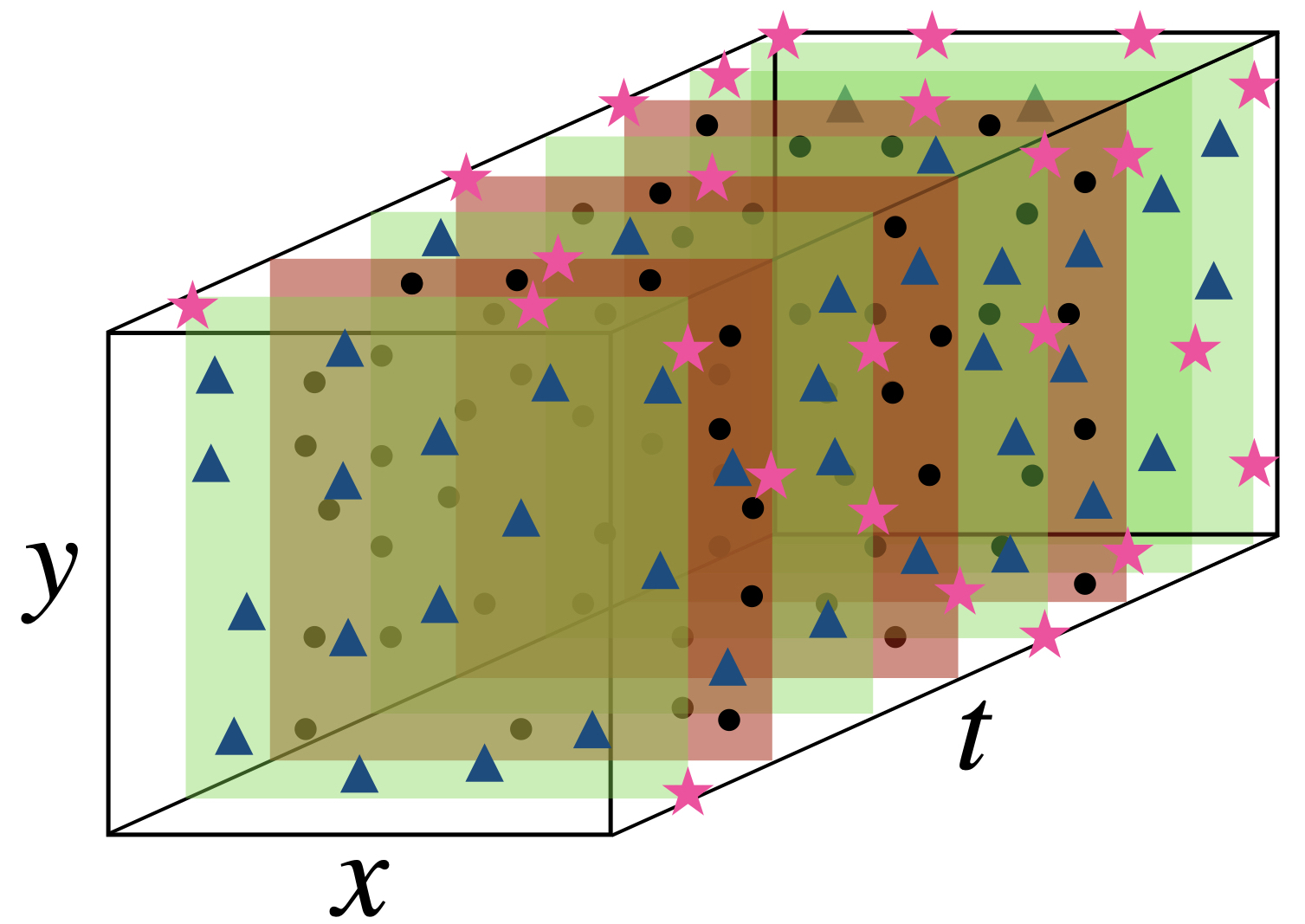}
\caption{\label{fig:schema} {\markup{Different sets of collocation points}} in the computational domain where equation losses (in ${\color{darkcerulean} \blacktriangle}$), data losses (in $\bullet$) and boundary losses (in ${\color{brightpink} \bigstar}$) are calculated. For the purpose of clarity, the collocation points corresponding to initial loss is not shown. }
\end{figure}

The inputs to the neural network are the spatial and temporal coordinates $({\bf x},t)$. According to \cite{goodfellow}, each network layer output is given as: 
\begin{align}
    {\bf z}_l = \sigma_l \left( {\bf z}_{l-1} {\bf W}_l + {\bf b}_l \right), 
\end{align}

where $l$ is the layer number, ${\bf W}$ represents the weight, ${\bf b}$ the bias and $\sigma_l$ is the activation function. The weights and biases are optimized so as to minimize the total loss $L_{tot}$. A gradient-based optimizer is used to perform the optimization \citep{gradient}, namely \textit{Adam} optimizer \citep{adam} is used for the first set of iterations and then we use L-BFGS-B \citep{lbfgs} for the remaining iterations until the set tolerance value is reached. 
\subsection{\label{sec:Turb_intro} Turbulence in periodic box}

We study forced dissipative turbulence in a doubly periodic domain of dimensions $ (x, y) \in [0, L] \times [0, L]$. The governing equation written in terms of the streamfunction $\psi$ is given by,
\begin{align}
    \partial_t \Delta \psi + J (\Delta \psi, \psi) = \nu \Delta^2 \psi - \alpha \Delta \psi + \Delta f_{\psi}, \label{eq:NS}
\end{align}
where $f_{\psi}$ is the forcing which drives the fluid motion and $\nu$ is the viscosity while $\alpha$ is the large scale friction.  $J(\Delta \psi, \psi)$ denotes the Jacobian given by $J(f,g) = \partial_x f \partial_y g - \partial_x g \partial_y f$. The forcing is chosen to be the Kolmogorov forcing $f_{\psi} = f_0 \cos \left( k_f y \right)$ where $k_f$ is the forcing wavenumber. Eq. \ref{eq:NS} with a spatial forcing, viscous dissipation and a large scale damping term is typically used to study two-dimensional turbulence \citep{sivashinsky1985weak, boffetta2012two, dallas2020transitions,seshasayanan2021bifurcations}. 

The velocity field and the vorticity field can then be calculated from the streamfunction using ${\bf u} = {\bm \nabla} \times \left( {\psi {\bf e}_z } \right)$, $\omega = - \Delta \psi$ respectively. The RMS velocity is denoted by $U_{rms}$ and is defined as $U_{rms} = \sqrt{ \left\langle {\bf u}^2\left( {\bf x}, t \right) \right\rangle_{{\bf x}, t}  }$ where $\left\langle \cdot \right\rangle_{{\bf x}, t }$ denotes the spatial and temporal averaging. The Reynolds number is defined using the RMS velocity $U_{rms}$, the length scale $ L$ and the viscosity $\nu$ as $Re = U_{rms} L/\nu$. Similarly for the large scale friction parameter $\alpha$, we can define a large scale Reynolds number $Rh = U_{rms}/\left( \alpha L \right)$. Two-dimensional turbulence shows an inverse cascade of energy and a forward cascade of enstrophy \citep{Kraichnan67}. Thus we need to resolve both the scales above the forcing scale and the scales below the forcing scale. We will look at both global and local quantities to study the statistical properties of turbulence. The energy $E (t) = U^2_{rms}(t)$ and the enstrophy $\mathcal{E} (t) = \left\langle \left( {\bm \nabla} \times \left( \psi \left( {\bf x}, t \right) {\bf e}_z \right) \right)^2 \right\rangle_{{\bf x} }$ are conserved in the absence of dissipation and forcing. The two-dimensional Fourier transform of the velocity field ${\bf u} \left( {\bf x}, t \right)$ is defined as,
\begin{align}
    \hat{\bf u} \left( {\bf k}, t \right) = \sum_{x_n=0}^L \sum_{y_n=0}^L {\bf u} ({\bf x}_n, t) \exp \left( -i {\bf k} \cdot {\bf x}_n \right), \label{eq:Fourier}
\end{align}
with ${\bf k} = (k_x, k_y)$ denoting the wavenumber vector. We then define the Energy spectra denoted by $E(k, t)$ as,
\begin{align}
    E(k, t) = \sum_{|{\bf k}| = k} |\hat{\bf u} ({\bf k}, t)|^2. \label{eq:Spectra_def}
\end{align}

\subsection{\label{sec:Turb_simul} Pseudospectral approximation}

The equations are solved via a pseudo-spectral method using Fourier basis functions, see Eq. \eqref{eq:Fourier}. Time-marching is done using the four step third order Runge Kutta scheme (ARS443), \citep{ascher1997implicit} and the aliasing errors are
removed with the two-third dealiasing rule. The domain is discretized into $N_x \times N_y$ points with $N_x$ along the $x$ direction and $N_y$ along the y direction. The pseudo-spectral solver \markup{based on \cite{gomez2005parallel}} is written in CUDA-based python code which is run on a GPU card. The resolution of the simulation is chosen based on the Reynolds number of the flow, for the maximum Reynolds number of $4000$ we have used a resolution of $512^2$. 

%
%
%

\section{PINN based models}

\begin{figure*}
\includegraphics[width=\textwidth,height=\textheight,keepaspectratio]{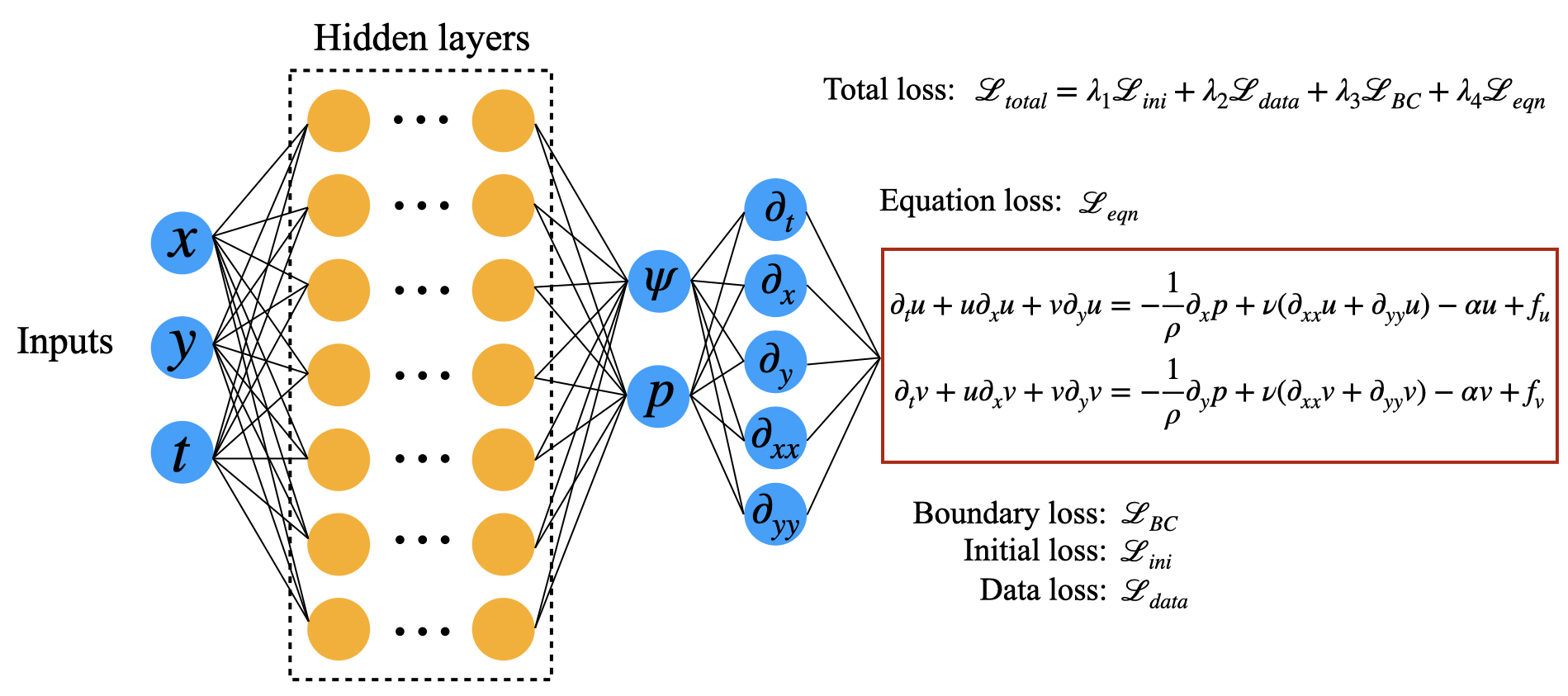}
\caption{\label{pinn1} Physics-informed neural network layout for PINN-1.}
\end{figure*}

\begin{figure*}
\includegraphics[width=\textwidth,height=\textheight,keepaspectratio]{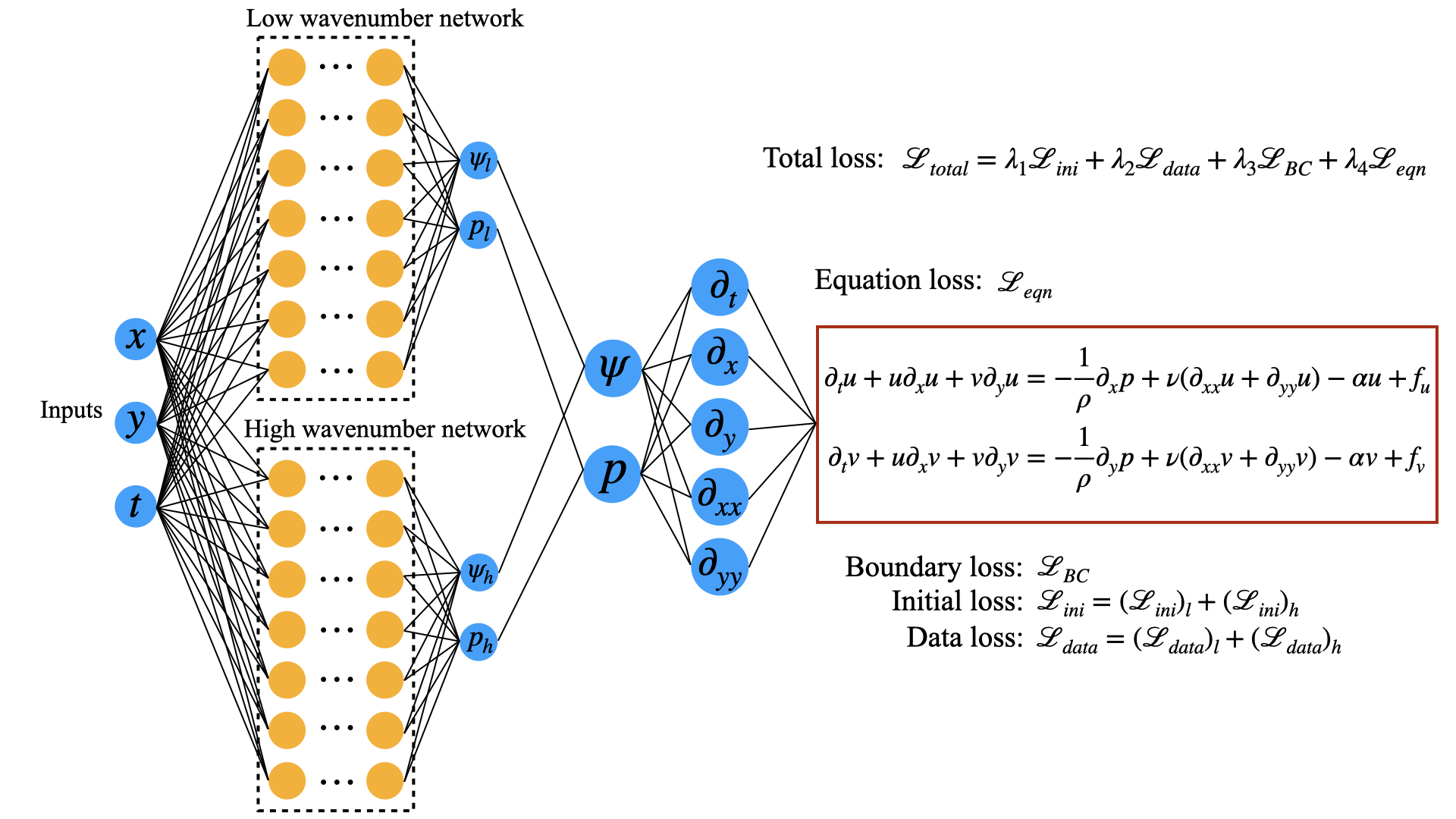}
\caption{\label{pinn2} Physics-informed neural network layout for PINN-2. The underscore letters "l" and "h" represent low and high wavenumbers.}
\end{figure*}

We construct two different PINN based models for our study. In both the approaches, we use the standard PINN model and on top of that we use some training data from the interior of the domain $(x,y)$ at certain time intervals \citep{lucor22}. We will refer to the first neural network as PINN-1 and it’s layout is shown in Fig. \ref{pinn1}. The output from the neural network is the predicted time evolving streamfunction and pressure fields which automatically satisfies the incompressibility condition. As for the second approach, we construct the neural network based on our understanding of the energy spectrum of turbulence. We know that the energy spectrum consists of a continuous distribution of energy across different length scales. We then construct neural networks for the low and high wavenumber parts where the training data which is fed to the network are from the low and high wavenumber components respectively. Previous study by \cite{ameya20} and \cite{xpinns} shows that splitting the training data in the physical domain can lower the value of overall loss function and is able to better predict the system. Motivated by their study, we do a decomposition in the spectral domain of the training data into low and high wavenumber components. It is known that a simple co-ordinate based selection of points leads to a good prediction of low-wavenumber behaviour while the high wavenumber modes are not captured \citep{tancik20}. We expect that the high wavenumber part of the flow will be better captured by this neural network architecture of separate networks for low and high wavenumber components.

The neural network layout for this approach is shown in Fig. \ref{pinn2} and we refer to it as PINN-2.  {\markup{As shown in }}Fig. \ref{pinn2}, the inputs for the pair of low and high wavenumber neural networks is the same set of sampling points $(x,y,t)$. The outputs from each of these neural networks are the predicted streamfunction and pressure fields corresponding to the low and high wavenumber parts. These outputs are further combined into the total streamfunction and pressure fields. These combined fields are then used for training on the equation and boundary conditions. {\markup{However, initial and data losses are separately calculated for the low and high wavenumbers}} and added together to give the final $\Lossini$ and $\Lossdata$. The data and initial fields that are fed into the two neural networks (low and high) {\markup{do not have}} any overlap in the spectral space but the outputs from these neural networks have an overlap in the spectral space due to the nonlinearity of the neural networks. The cutoff wavenumber $k_c$ for the PINN-2 model is a parameter which separates the low and high wavenumber part of the flow field. 
The total loss for the model with low and high wavenumber networks will have the same expression as Eq. $\eqref{eqn:Loss_def}$, except that the individual losses $\Lossini$ and $\Lossdata$ are defined for PINN-2 as $\Lossini = (\Lossini)_l + (\Lossini)_h$ and $\Lossdata = (\Lossdata)_l + (\Lossdata)_h$. 
Here $(\Lossini)_l$ represents the initial loss calculated for the low-wavenumber network while $(\Lossini)_h$ represents the initial loss for the high-wavenumber network, their definitions remain the same as the one for $\Lossini$, see Eq. \eqref{eqn:loss_ini}. 
Similarly $(\Lossdata)_l,(\Lossdata)_h $ represent the data loss calculated for the low and high wavenumber networks respectively. 
Their definitions remain the same as the one for $\Lossdata$, see Eq. \eqref{eqn:loss_data}. One can in-principle extend this idea to construct multiple cutoff wave-numbers, for example, we can construct high, mid and low-wavenumber network models. We leave such extensions for future studies.

We see that the definitions of the Loss function for the two PINN networks, PINN-1 and PINN-2 are different since loss initial $\Lossini$ and loss data $\Lossdata$ are calculated twice for PINN-2. 
It should be noted that the structure of the neural network in PINN-2 can support different number of neurons, layers for the low and high wave number networks. 
Since the number of weight parameters and biases depend on the number of neurons and layers, we can thus control the number of these parameters for low and high wavenumber networks independently.  

\section{Results and discussion}

We fix the forcing amplitude $f_0 = 1$ and the forcing wavenumber to be $k_f L = 8 \pi$. Starting from an arbitrary initial condition the forced two-dimensional turbulence leads to a statistical stationary state after an initial transience. We choose a set of $\nu, \alpha$ that leads to a Reynolds number of $Re \approx 4.2 \times 10^3$ and a large scale Reynolds number of $Rh \approx 2.1$ which is used as a database for the entire study. We will focus our attention on this statistical stationary state of turbulence to develop our PINN based model. The aim of the model is to capture the statistical properties of the turbulent flow over a fixed time interval for different temporal and spatial scales. We will first identify the best performing neural network and then proceed to compare it with DNS. 

\subsection{Results from PINNs: training \& hyperparameter search}

We define the hyperparameters, see in Table \ref{tab:hyperparameters}, over which different neural networks are trained. In the first row of the table, the number of neurons for the PINN-2 model is defined as $(n_{\text{low}}, n_{\text{high}})$ where $n_{\text{low}}$ is the number of neurons in the low wavenumber network and $n_{\text{high}}$ is the number of neurons in the high wavenumbers network.  
\begin{table*}
\caption{\label{tab:table1}Hyperparameters}
\begin{ruledtabular}
\begin{tabular}{lcc}
Parameters & PINN-1 & PINN-2 \\
\hline
Num. neurons & 250,\ 300,\ 350, 400 & (100,150),\ (50,250),\ (100,250),\ (250,250),\ (100,350) \ (low,high)\\
Num. layers & 5,\ 7 & 5,\ 7\\
Num. training data & 0.1\%,\ 0.2\%,\ 0.4\% & 0.1\%,\ 0.2\%,\ 0.4\%
\end{tabular}
\label{tab:hyperparameters}
\end{ruledtabular}
\end{table*}
Losses from equation and boundary are minimized at $81$ instances in the time interval $t \in [0,T]$ with equally spaced time-steps of $T/80$. 
The data losses are minimized only from $21$ equally spaced snapshots with time steps of $T/20$, while the initial loss is computed from $t = 0$. 
We will denote these $21$  time instances where the data or initial condition is provided as training times. 
The rest $60$ time instances are denoted as testing times since interior data is not provided to the PINN model and we plan to compare the performance of the PINN with the DNS data. 
For equation losses we have used $N_{eqn} = 10000$ points which are randomly selected from $512 \times 512 \times 81$ points.
For initial points, we have chosen $N_{ini} = 1000$ points out of $512 \times 512$ points at the time instant $t = 0$. 
The boundary points $N_{BC} = 2000$ are selected randomly from $4 \times 512 \times 81$ points and the data points $N_{data} = 5000$ are selected randomly from $512 \times 512 \times 21$ points. The data points $N_{data}$ {\markup{correspond}} close to $0.1\%$ of the total available data points.  
Filtering is done to separate the energy into low and high wavenumbers. 
The cut-off wavenumber is taken to be slightly above the forcing wavenumber at $k_c L = 20 \pi$. {\markup{We have also studied the results with $k_c L = 40 \pi$ and $k_c L = 60 \pi$ and found that $k_c L = 20 \pi$ gives the lowest error estimates, the definitions of which will be introduced in the next sub-section}}. 

\begin{figure*}
\begin{subfigure}[t]{0.495\textwidth}
\includegraphics[width=\textwidth,height=\textheight,keepaspectratio]{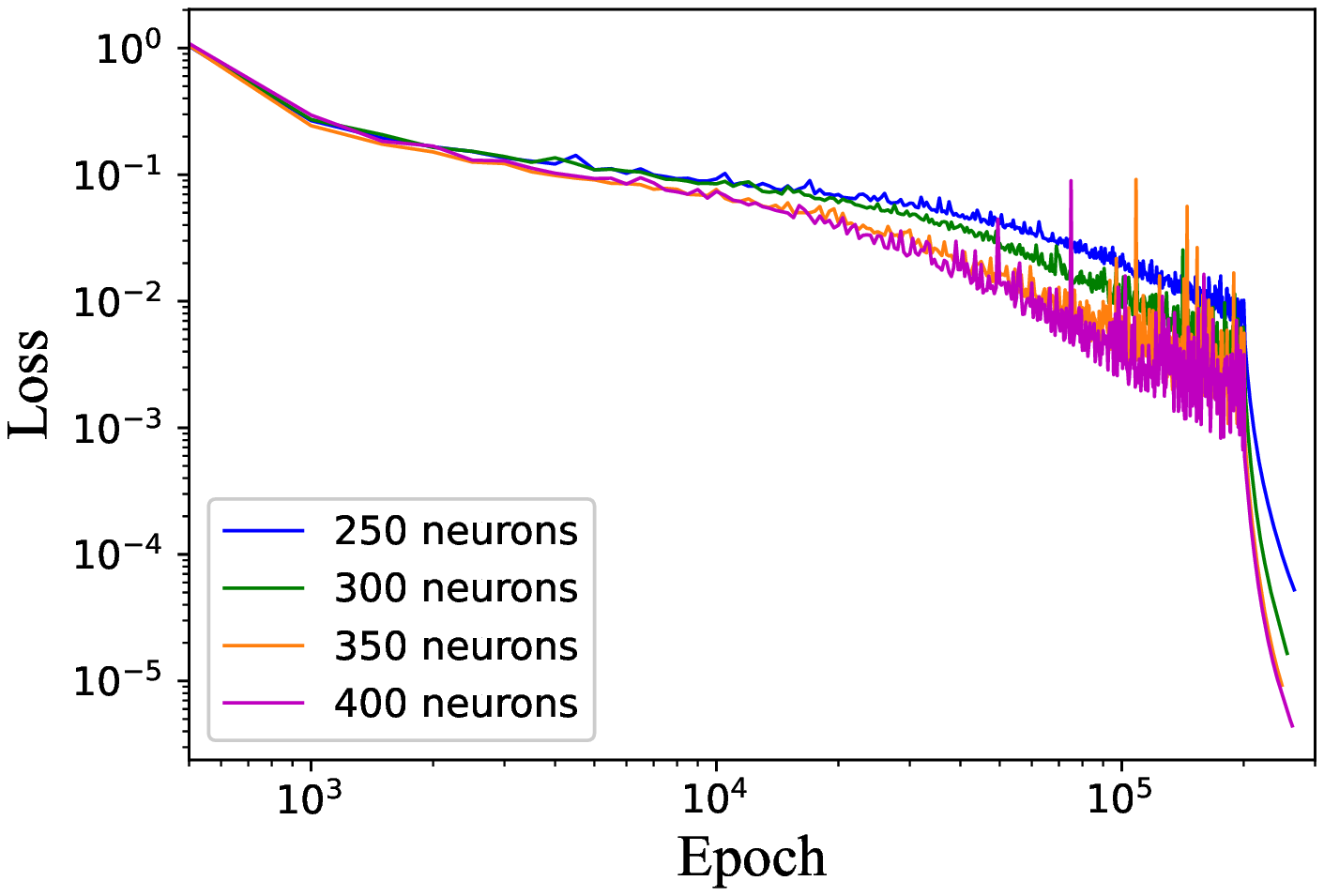}
\caption{}\label{fig:loss_neurons_PINN1}
\end{subfigure}
\begin{subfigure}[t]{0.495\textwidth}
\includegraphics[width=\textwidth,height=\textheight,keepaspectratio]{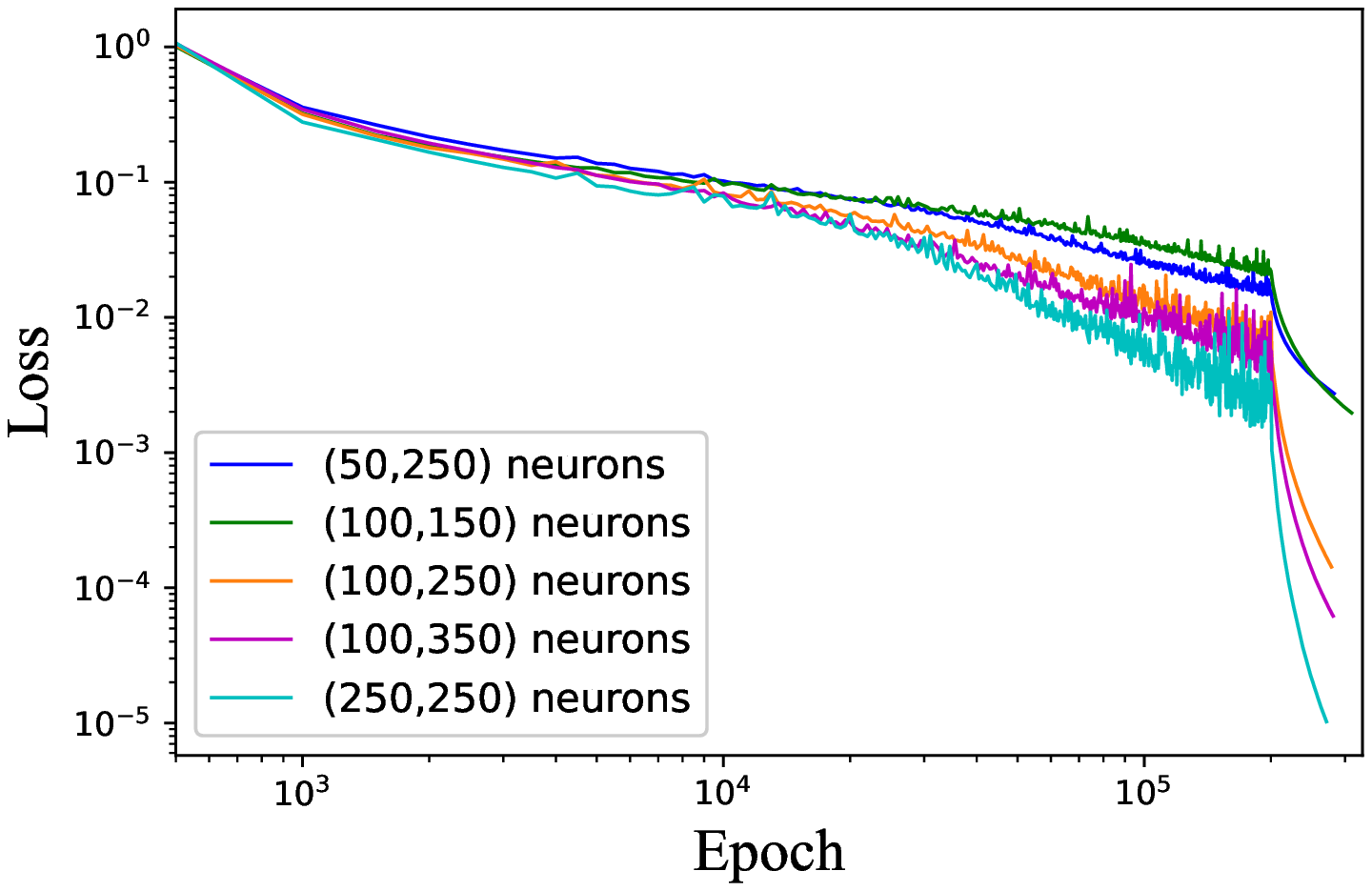}
\caption{}\label{fig:loss_neurons_PINN2}
\end{subfigure}
\caption{\label{fig:wide} {\markup{Total loss}} for different number of neurons for (a) PINN-1 model and (b) PINN-2 model.}
\label{fig:loss_neurons}
\end{figure*}
\begin{figure*}
\begin{subfigure}[t]{0.495\textwidth}
\includegraphics[width=\textwidth,height=\textheight,keepaspectratio]{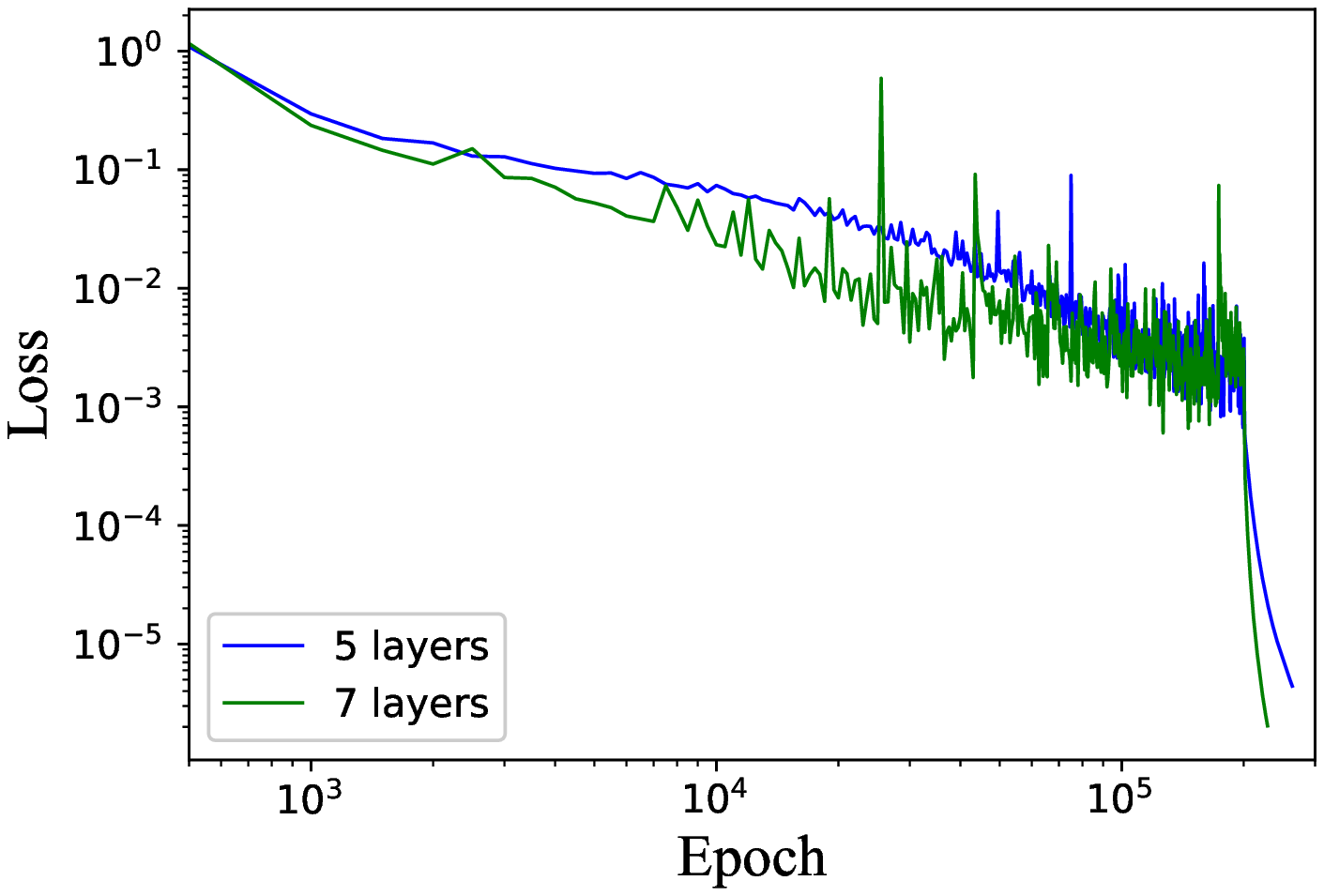}
\caption{}\label{fig:loss_layers_PINN1}
\end{subfigure}
\begin{subfigure}[t]{0.495\textwidth}
\includegraphics[width=\textwidth,height=\textheight,keepaspectratio]{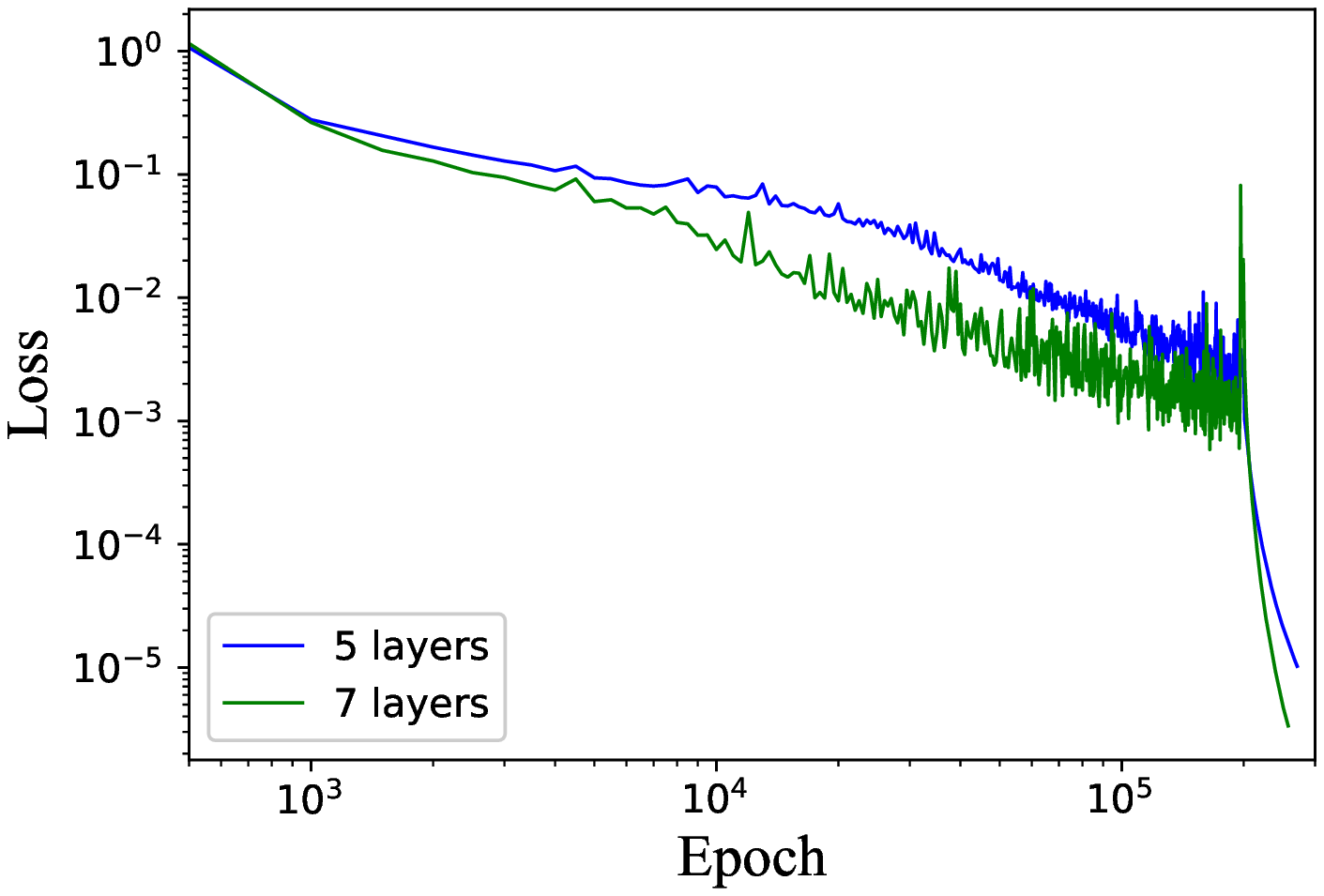}
\caption{}\label{fig:loss_layers_PINN2}
\end{subfigure}
\caption{\label{fig:wide} {\markup{Total loss}} for different number of layers for (a) PINN-1 model and (b) PINN-2 model.}
\label{fig:loss_layers}
\end{figure*}
\begin{figure*}
\begin{subfigure}[t]{0.495\textwidth}
\includegraphics[width=\textwidth,height=\textheight,keepaspectratio]{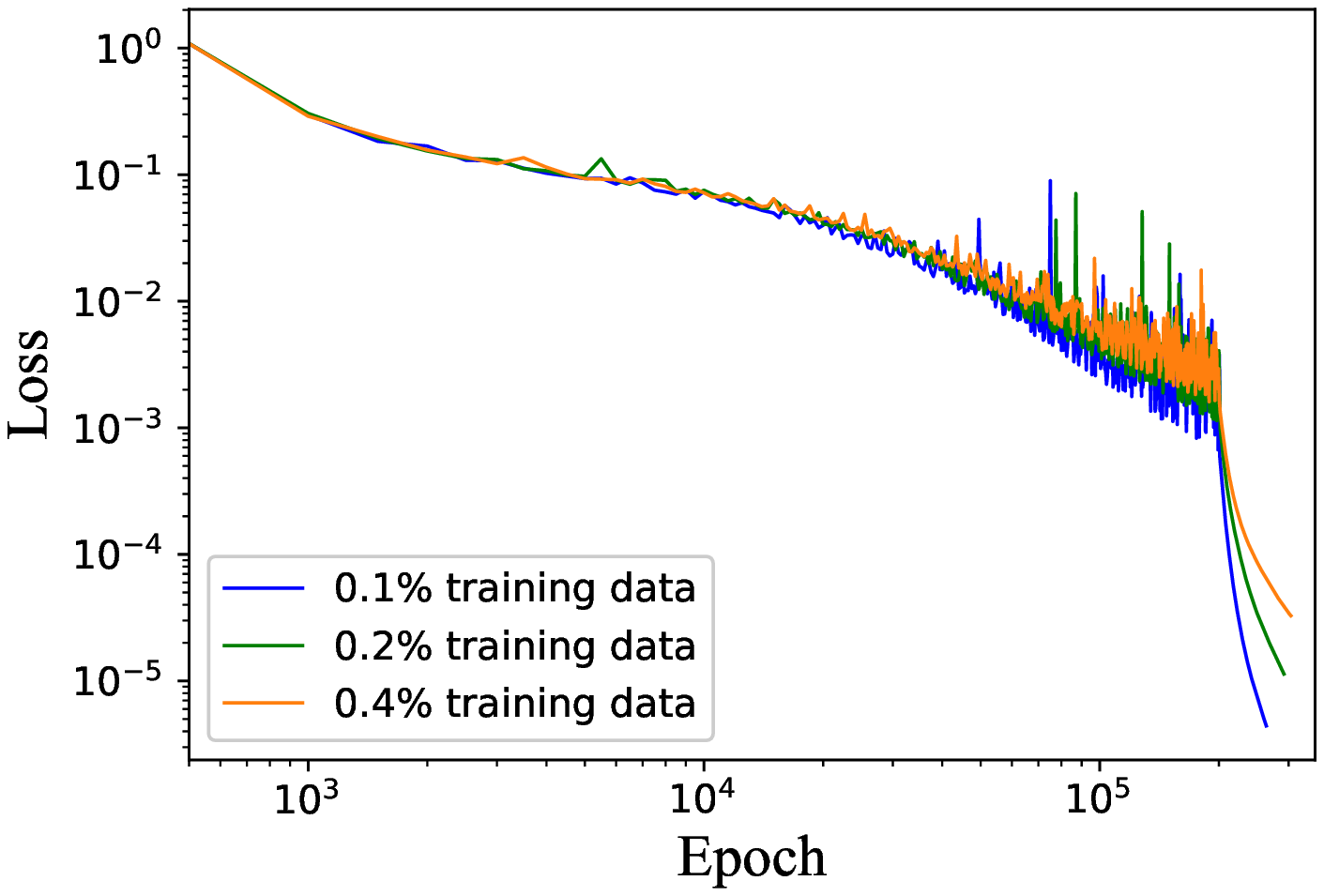}
\caption{}\label{fig:loss_data_PINN1}
\end{subfigure}
\begin{subfigure}[t]{0.495\textwidth}
\includegraphics[width=\textwidth,height=\textheight,keepaspectratio]{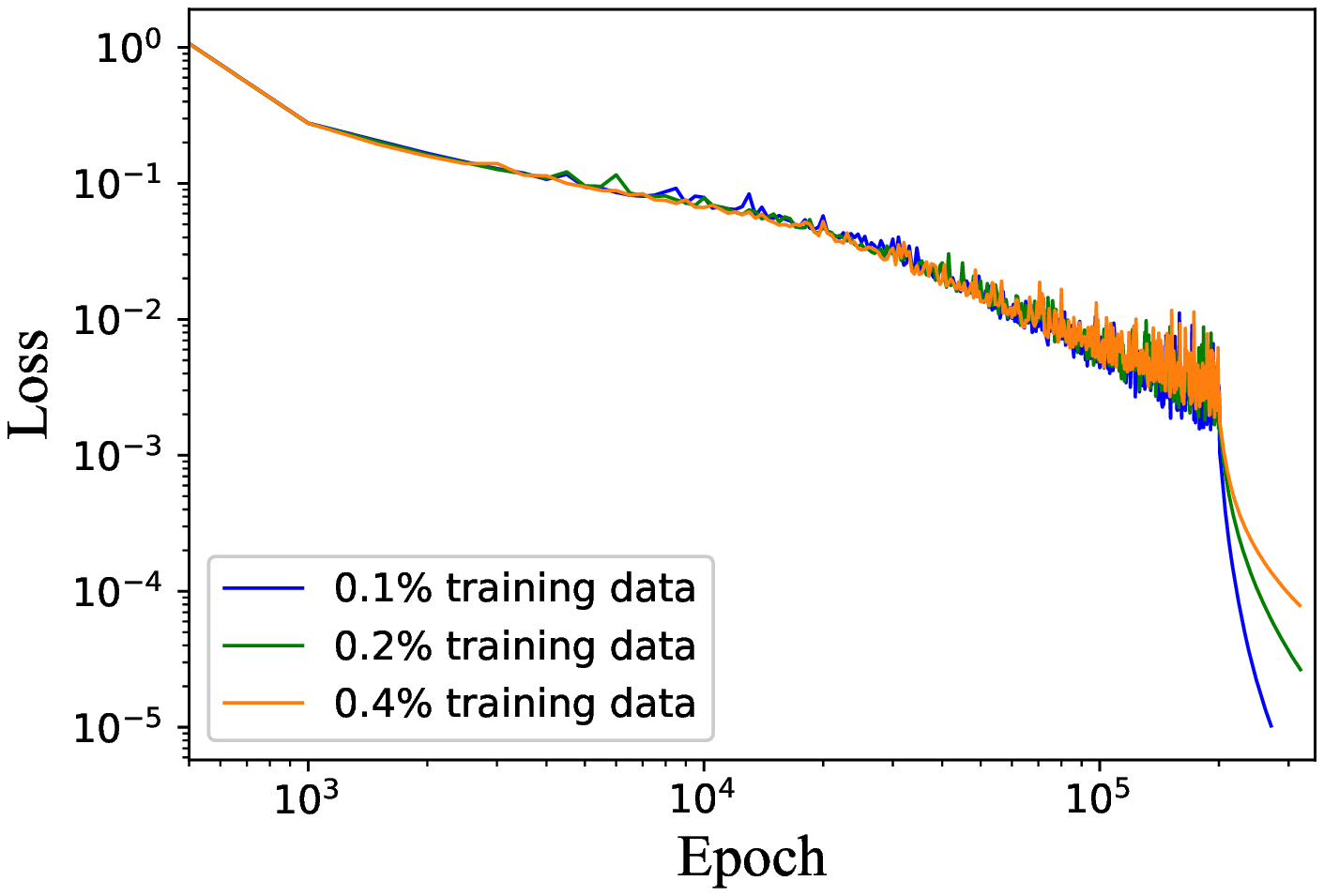}
\caption{}\label{fig:loss_data_PINN2}
\end{subfigure}
\caption{\label{fig:wide} {\markup{Total loss}} for different number of training data for (a) PINN-1 model and (b) PINN-2 model.}
\label{fig:loss_data}
\end{figure*}
For all our hyperparameter sweeps, we train the model for $200000$ {\markup{epochs (iterations)}} for the \textit{Adams} optimizer and then we use upto $200000$ epochs using L-BFGS-B optimizer until the tolerance reaches machine precision (double precision). 
We use a learning rate $\eta=0.001$ for the \textit{Adams} iterations. 
In this study, we use a $\tanh$ activation function for all the layers except for the last layer where we use a linear activation function \citep{raissi19}. 
The training is performed using a NVIDIA Tesla V100 card. First, we do a sweep of the list of neurons listed in Table \ref{tab:hyperparameters} for both PINN-1 and PINN-2 models. As shown in Fig. \ref{fig:loss_neurons_PINN1}, we observe for PINN-1 model that the total loss value for $400$ neurons was the lowest among others. On the other hand, for the PINN-2 model shown in Fig. \ref{fig:loss_neurons_PINN2}, the combination of (250,250) neurons for low and high wavenumber networks respectively, resulted in the least total loss value. We also remark that this PINN-2 model has $2 \times \left( (3 \times 250) + 2 \times (250 \times 250) + (2 \times 250) \right) = 252500$ trainable weight parameters. 
In comparison, PINN-1 model with $400$ neurons has $(3 \times 400) + 2 \times (400 \times 400) + (2 \times 400) = 322000$ trainable weight parameters. 
Even though the trainable weight parameters in this case is more than $25\%$ higher than {\markup{the}} best PINN-2 model, going forward we will find that the PINN-2 model performs better than the PINN-1 model. 

We now take the best performing set of neurons for the PINN-1 and PINN-2 models and perform the next hyperparameter search for the optimal number of layers as listed in Table \ref{tab:hyperparameters}. 
For the PINN-1 model, we show in Fig. \ref{fig:loss_layers_PINN1} the loss functions for the sets of layers as shown in the Table \ref{tab:hyperparameters}. 
We see that the layer $5$ network does almost as good as the layer $7$ network, so we continue to use $5$ layer model for further exploring PINN-1. 
Similarly from Fig. \ref{fig:loss_layers_PINN2} the $5$ layer model performs as well as the $7$ layer model. Thus we continue with the $5$ layer model for both PINN-1 and PINN-2. 
Next, we find the loss function for different percentages of training data.
We reiterate that the training data is taken from the interior of the simulation domain from the existing DNS solution. 
In Figs. \ref{fig:loss_data_PINN1}, \ref{fig:loss_data_PINN2} we show the losses for different percentage of training data for both PINN-1 and PINN-2 networks. {\markup{As we increase the percentage of training data we see that the total loss increases marginally because the $\Lossdata$ increases. This behavior is observed generally in machine learning models \citep{andrewng_book}. We will continue with $0.1\%$ percentage for both the PINN-1 and PINN-2 because it has comparable total loss values with the other training data sets and that it is computationally cheaper}}.


\subsection{Comparison with DNS results}

{\markup{One of the main aim of this work is to accurately capture the modes using the proposed PINN model. We first observe the energy spectra $E(k)$ which gives the energy content across different length scales}}.
We define the quantity $\epsilon_{Ek} = \sum_k |\log \left( \overline{E}_{\text{PINN}} (k) /\overline{E}_{\text{DNS}} (k)\right) |$ which denotes the deviation of the mean testing time energy spectra from PINN model ${\overline{E}_{\text{PINN}}}$ with that of the DNS $\overline{E}_{\text{DNS}}$. {\markup{Here $\overline{E} (k)$ denotes the mean of the energy spectra found at testing times. The spectra are averaged over $60$ testing time instances}}.
{\markup{To find the best PINN model, we have tried a variety of values for $\lambda_4 \in (1,50,75,100,150,200)$. We show in Table \ref{tab:table2} the values of $\epsilon_{Ek}$ for different $\lambda_4$ values for both PINN-1 and PINN-2 models.  We find $\lambda_4 = 150$ for the PINN-1 model and $\lambda_4 = 75$ for the PINN-2 model to perform best, since they have the lowest $\epsilon_{Ek}$ among all the $\lambda_4$'s examined. This is also seen from Fig. \ref{fig:lambda4_PINN1} which shows the energy spectra $\overline{E}(k)$ for different values of $\lambda_4$ for the PINN-1 model while Fig.  \ref{fig:lambda4_PINN2} shows $\overline{E} (k)$ for different values of $\lambda_4$ for the PINN-2 model.}
\begin{figure*}
\begin{subfigure}[t]{0.495\textwidth}
\includegraphics[width=\textwidth,height=\textheight,keepaspectratio]{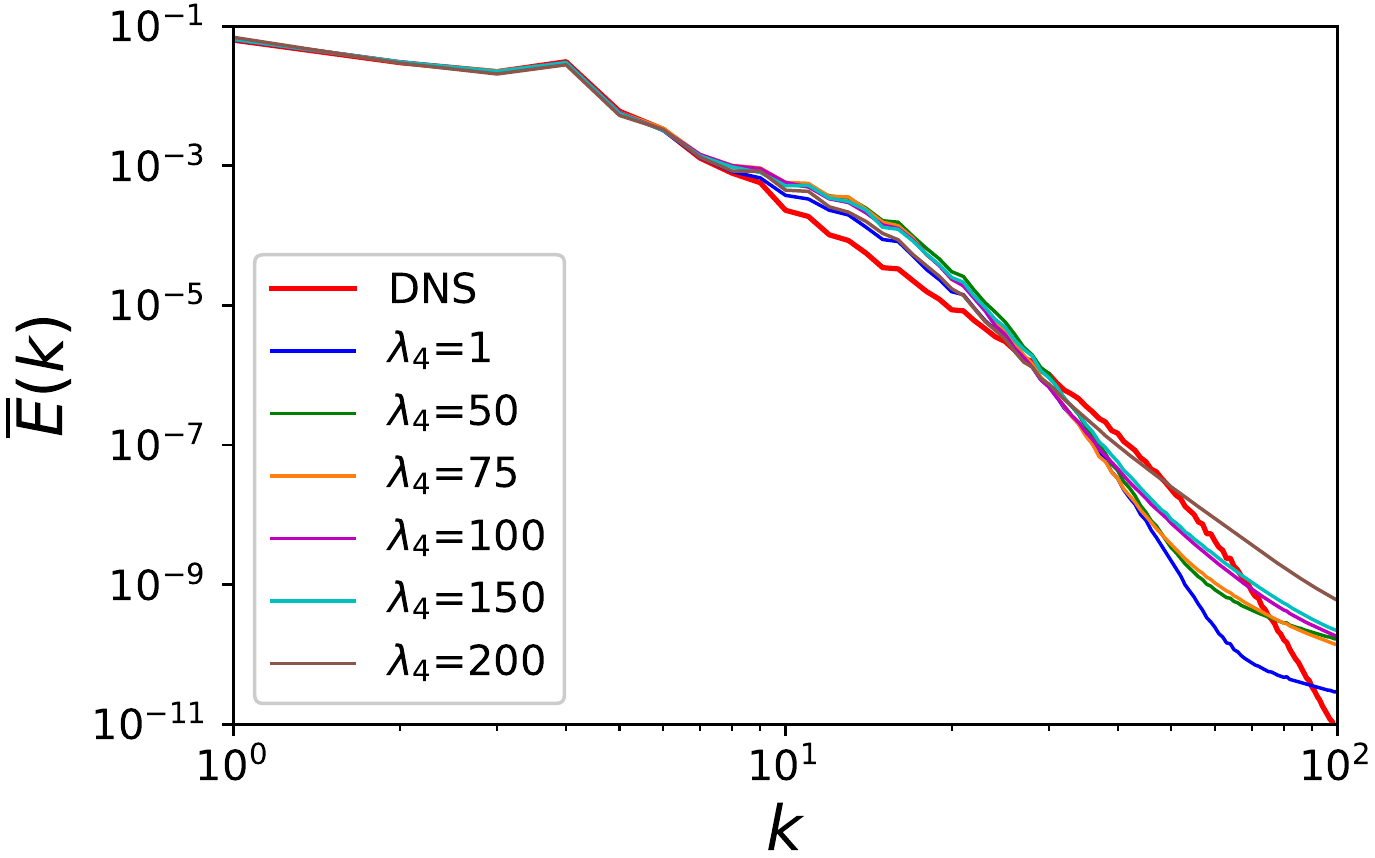}
\caption{}\label{fig:lambda4_PINN1}
\end{subfigure}
\begin{subfigure}[t]{0.495\textwidth}
\includegraphics[width=\textwidth,height=\textheight,keepaspectratio]{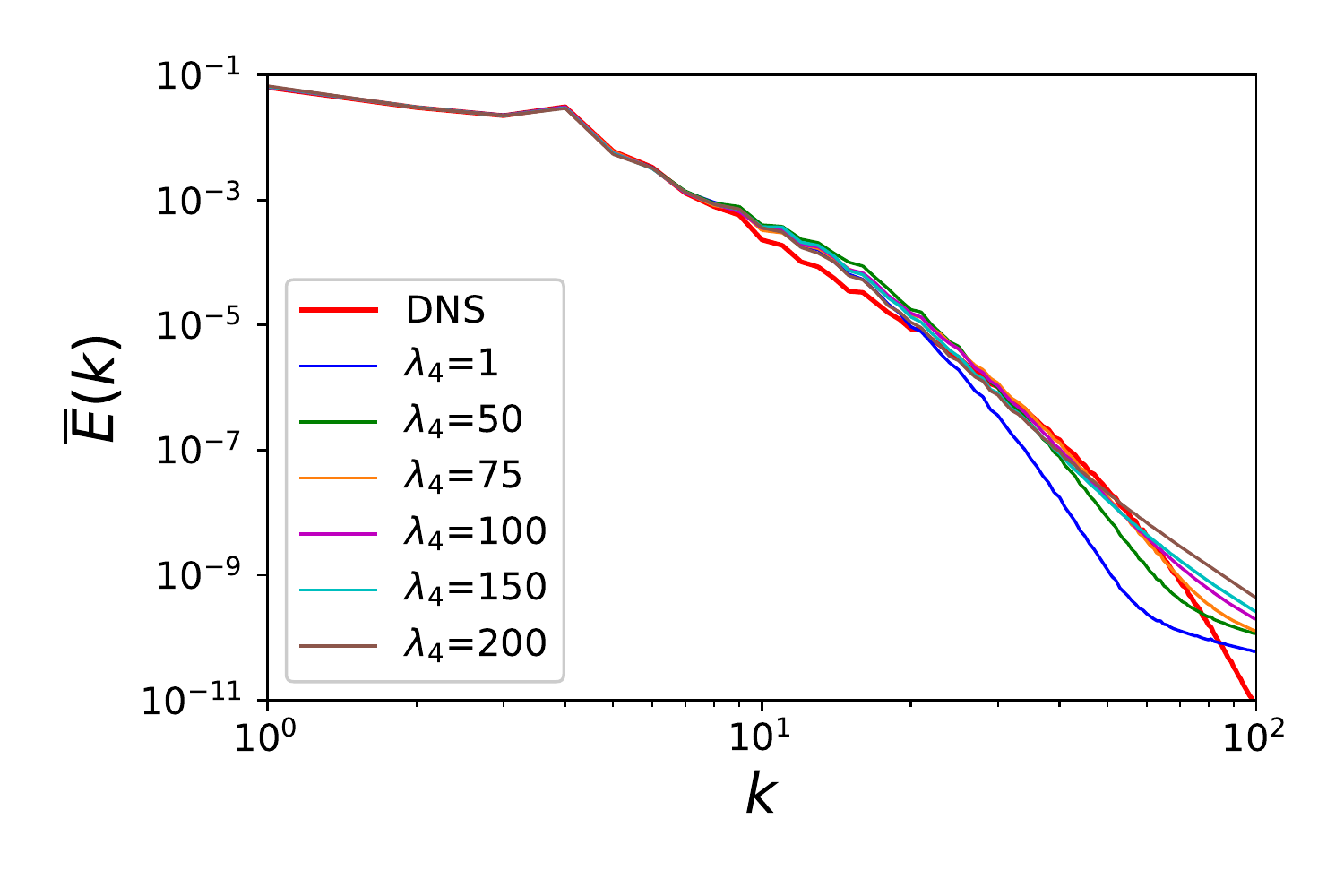}
\caption{}\label{fig:lambda4_PINN2}
\end{subfigure}
\caption{\label{fig:lambda4_spectra} {\markup{Plots of the energy spectra from {(a)} PINN-1 and {(b)} PINN-2 models for different values of $\lambda_4$ in comparison with DNS}}.}
\end{figure*}
Thus, for the weights in the expression Eq. $\eqref{eqn:Loss_def}$, we have set the values to be $\lambda_1 = \lambda_2 = \lambda_3 = 1, \lambda_4 = 150$ for the PINN-1 model and $\lambda_1 = \lambda_2 = \lambda_3 = 1, \lambda_4 = 75$ for the PINN-2 model. 
Large value of $\lambda_4$ implies that a larger significance is given to the equation loss so that we penalize the unphysical solutions that deviate from the governing equations. 
Also, we retrain the weights once and we observe better correlation of the energy spectra and energy time series with those from the DNS. Particularly, for the chosen value of $\lambda_4=150$ for PINN-1 and $\lambda_4=75$ for PINN-2 model, retraining once reduced the $\epsilon_{Ek}$ to $79.91$ and $47.2$ respectively. Further retraining does not reduce $\epsilon_{Ek}$ by more than $10\%$, so we only retrain the weights only once.

In what follows, we make all the quantities dimensionless by rescaling with the typical velocity scale $U\_$, length scale $L$ and the time scale given by $L/U$.

First, in order to find the deviation of our PINN based solutions from DNS, we define an error measure denoted as RMSE as follows,
\begin{align}
    \epsilon = \biggl( \frac{1}{n_t} \sum_{i}^{n_t} (\bf{u}_{\text{PINN}}^{i} - \bf{u}_{\text{DNS}}^{i})^2 \biggr)^{1/2},
    \label{eq:aRMSE}
\end{align}
where, $n_t = N^2 \times T = 512^2 \times 60$ is the total number of data points present at the testing times. 
We measure this error along with the coefficient of determination $R^2$ for the two components of the velocity. 
The RMSE is calculated over the whole domain because we aim to reproduce the whole field across all length scales. 
We tabulate these diagnostics for both PINN-1 and PINN-2 models in Table \ref{tab:table3}.
{\markup{
\begin{table}
\caption{\label{tab:table2}$\epsilon_{Ek}$ for PINN-1 and PINN-2}
\begin{ruledtabular}
\begin{tabular}{lcc}
$\lambda_4$ & $\epsilon_{Ek}$ (PINN-1) & $\epsilon_{Ek}$ (PINN-2) \\
\hline
1 & 125.65 & 136.49 \\
50 & 114.34 & 79.55 \\
75 & 110.02 & 55.51 \\
100 & 97.37 & 74.72 \\
150 & 97.05 & 87.26 \\
200 & 121.71 & 105.01 
\end{tabular}
\label{tab:errors}
\end{ruledtabular}
\end{table}
}
\begin{table}
\caption{\label{tab:table3}Errors for PINN-1 and PINN-2}
\begin{ruledtabular}
\begin{tabular}{lcccc}
Model & $\epsilon_{u}$ & $\epsilon_{v}$ & $(R^2)_{u}$ & $(R^2)_{v}$ \\
\hline
PINN-1 & 1.45e-1 & 1.40e-1 & 9.61e-1 & 9.58e-1 \\
PINN-2 & 1.16e-1 & 1.21e-1 & 9.75e-1 & 9.69e-1
\end{tabular}
\label{tab:errors}
\end{ruledtabular}
\end{table}
We clearly find that, for the parameters chosen to study PINN-1 and PINN-2, PINN-2 performs better with lower error measures and also has a higher $R^2$ score as compared to PINN-1. {\markup{The pressure field is also an output from the PINN models. For the PINN-2 model we find RMSE for pressure to be $0.09$}}.

\begin{figure*}
\begin{subfigure}[t]{0.495\textwidth}
\includegraphics[width=\textwidth,height=\textheight,keepaspectratio]{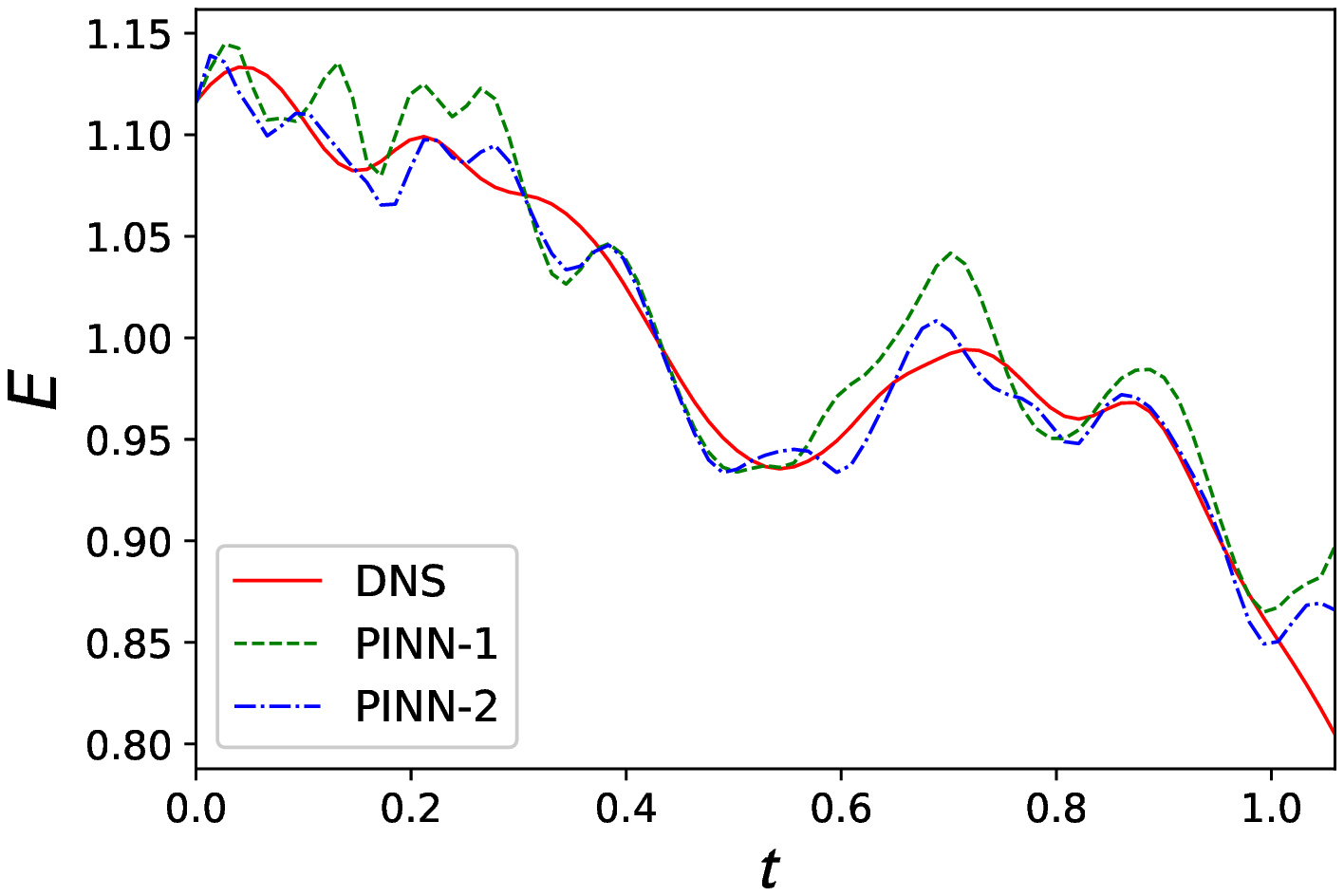}
\caption{}\label{fig:energy_series}
\end{subfigure}
\begin{subfigure}[t]{0.495\textwidth}
\includegraphics[width=\textwidth,height=\textheight,keepaspectratio]{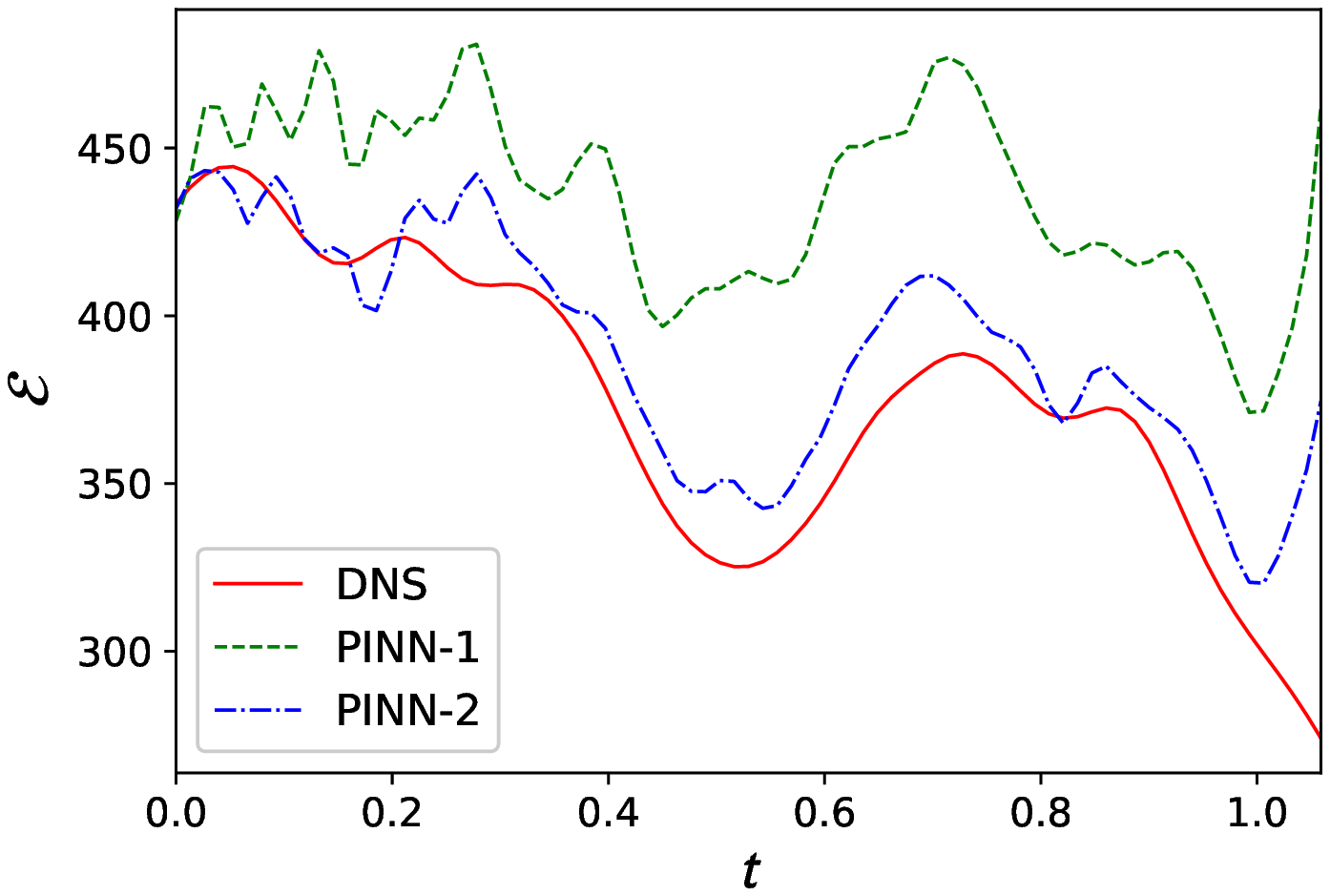}
\caption{}\label{fig:enstrophy_series}
\end{subfigure}
\caption{\label{fig:time_series} {\markup{(a)}} Plots of the time series of the energy from PINN-1 and PINN-2 models in comparison with DNS. {\markup{(b)}} Plots of the enstrophy from PINN-1 and PINN-2 models in comparison with DNS.}
\end{figure*}

Next, in Fig. \ref{fig:energy_series} we compare the time series of the total energy $E(t)$ of both the DNS and the PINN models. 
The PINN-2 model captures well the fluctuations of the total energy as a function of time as compared to the PINN-1 model. In Fig. \ref{fig:enstrophy_series} we compare the time series of the enstrophy between the DNS and the PINN models. PINN-1 model predicts a higher value of the enstrophy as compared with the DNS while PINN-2 performs better and compares well with the solution from DNS. 

\begin{figure}
\centering
\includegraphics[width=0.5\textwidth,height=0.5\textheight,keepaspectratio]{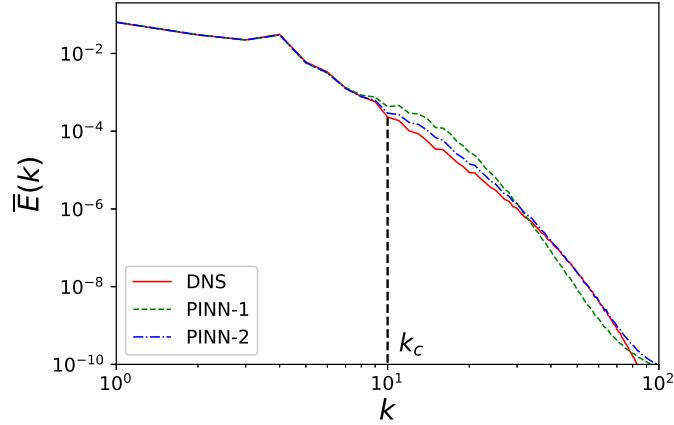}
\caption{\label{fig:energy_spectra} {\markup{Mean}} kinetic energy spectra $\overline{E}(k)$ from PINN-1 and PINN-2 models in comparison with DNS. {\markup{The dotted vertical line denotes the cutoff wavenumber $k_c$}}.}
\end{figure}


{\markup{Next, we look at the mean energy spectra $\overline{E}(k)$ for both PINN-1 and PINN-2 models}}. 
Fig. \ref{fig:energy_spectra} shows the mean energy spectra comparison between the DNS and the PINN models at testing times. The spectra from the PINN-2 model shows an excellent comparison with the DNS data. 
The PINN-2 spectra starts deviating from the DNS spectra at values of $E(k) \sim 10^{-9}$.
It is clear from the spectra that the large scales $k \lesssim 10$ agree to a very good degree at all times whereas intermediate and small scales start showing deviations around the DNS results.
The energy spectra is deviating at viscous scales due to the formation of very fine structures at the boundaries.
{\markup{To capture the energy spectra well into the viscous scales, one might have to enforce the periodic boundary conditions exactly \citep{dong2021method}. }}

We then focus on the snapshots of the velocity fields for the DNS and PINN-2 model in Fig. \ref{fig:contours_t1} and Fig. \ref{fig:contours_t2} for the testing times $t=0.45$ and $t=0.98$ respectively. {\markup{As seen from these figures, the large scale flow structures present in the velocity fields from the PINN-2 model are identical to the actual turbulent solutions from DNS even though we have used only $0.1 \%$ of the available DNS data for training. In the third column of the Figs. \ref{fig:contours_t1} and \ref{fig:contours_t2}, we show the absolute errors of the velocity fields to indicate deviations between the PINN-2 model and the DNS solutions}}. 
\begin{figure*}
\begin{subfigure}[t]{0.33\textwidth}
\includegraphics[width=\textwidth,height=\textheight,keepaspectratio]{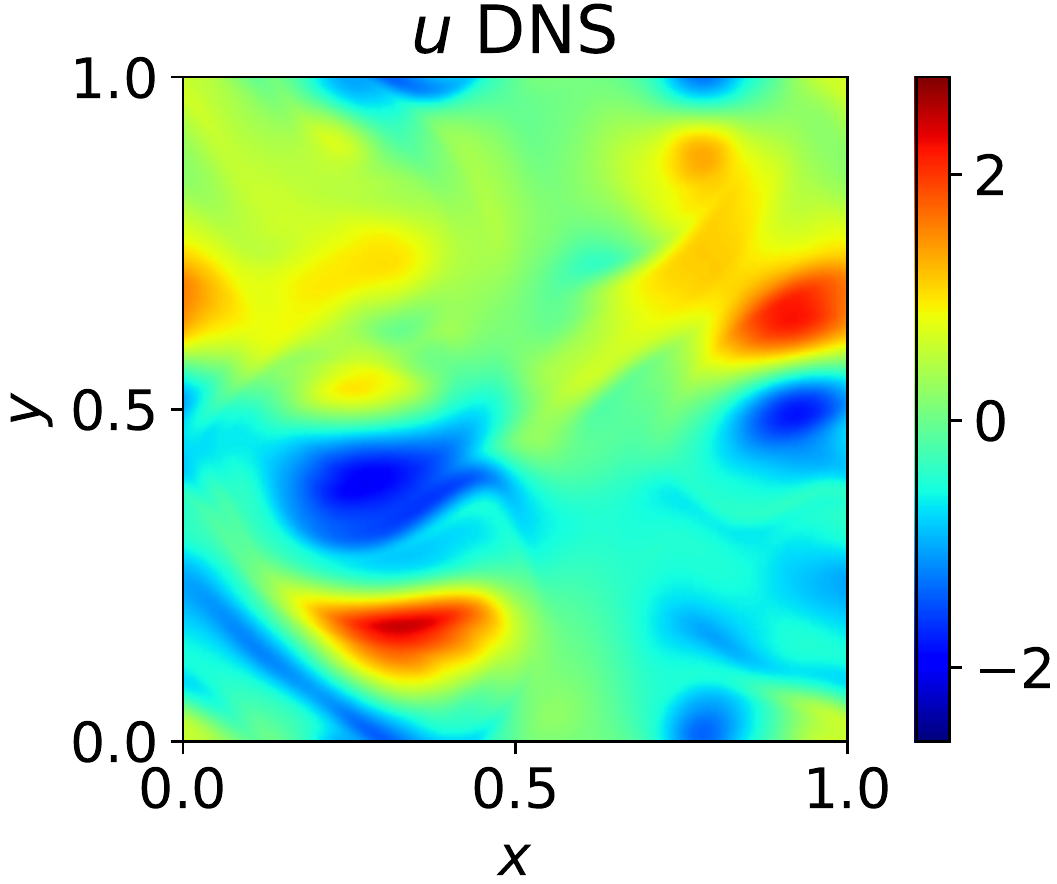}
\end{subfigure}
\begin{subfigure}[t]{0.33\textwidth}
\includegraphics[width=\textwidth,height=\textheight,keepaspectratio]{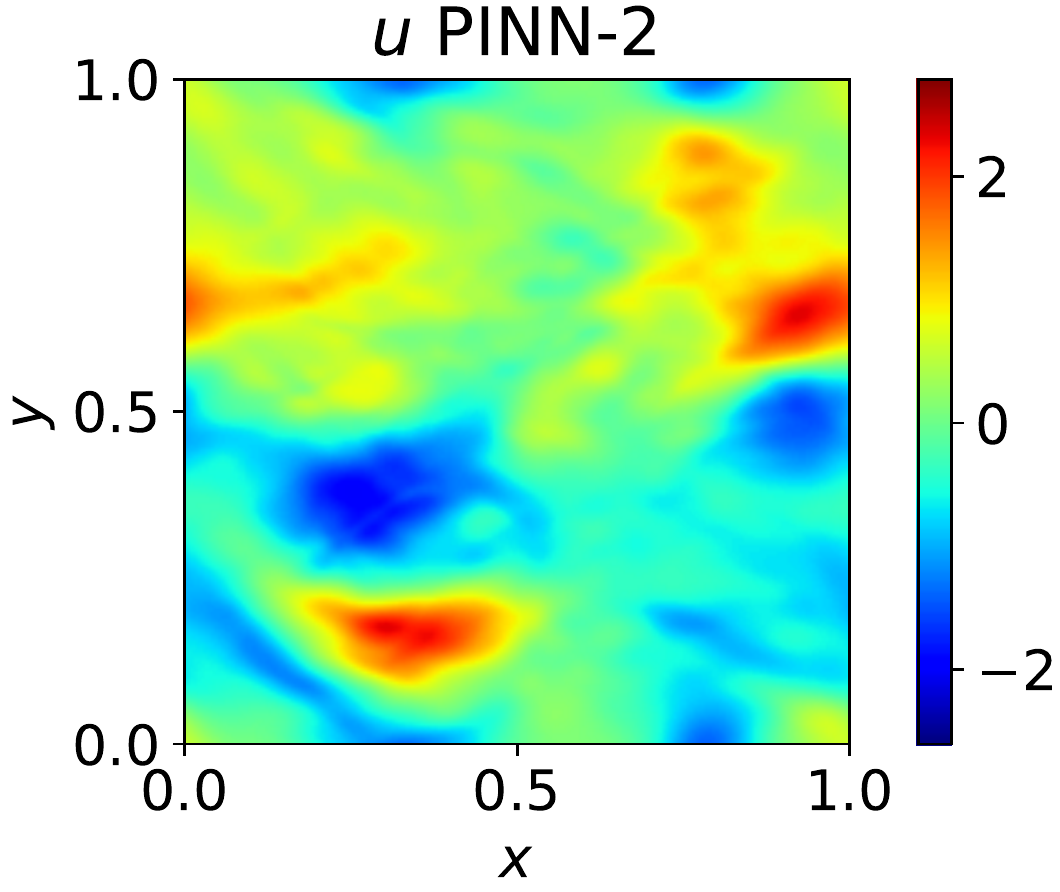}
\end{subfigure}
\begin{subfigure}[t]{0.33\textwidth}
\includegraphics[width=\textwidth,height=\textheight,keepaspectratio]{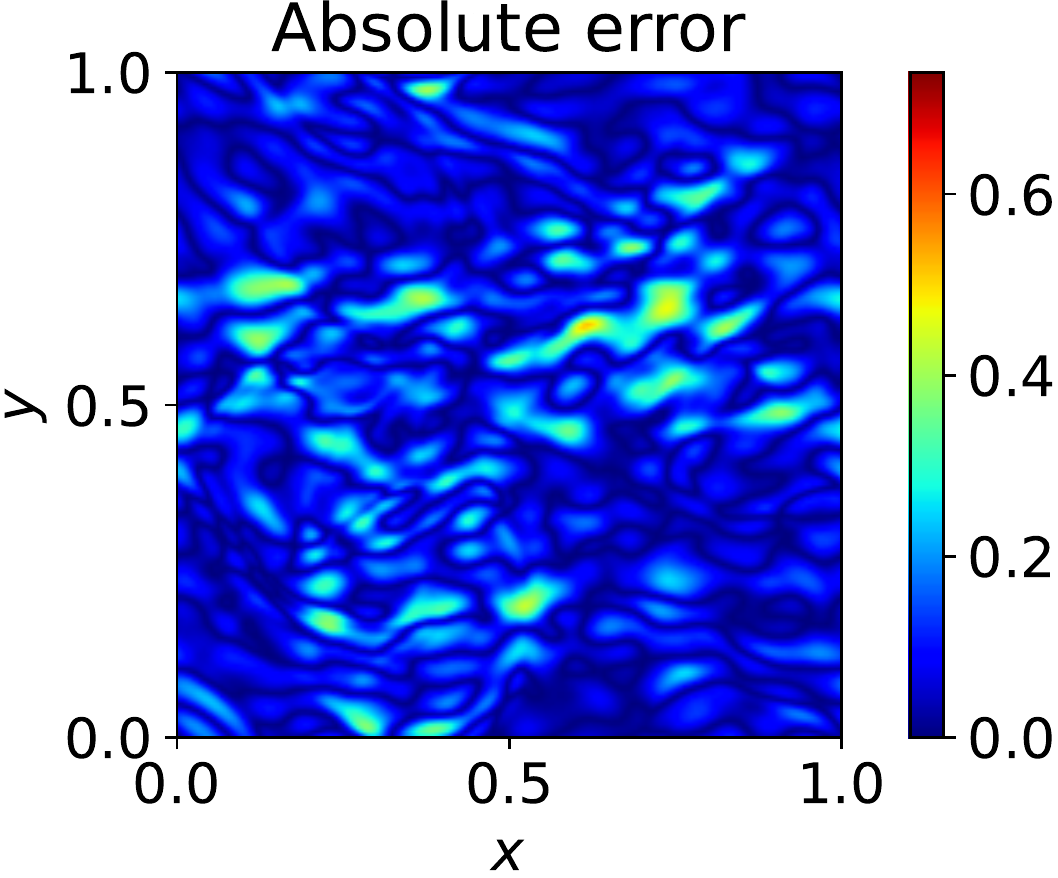}
\end{subfigure}
\begin{subfigure}[t]{0.33\textwidth}
\includegraphics[width=\textwidth,height=\textheight,keepaspectratio]{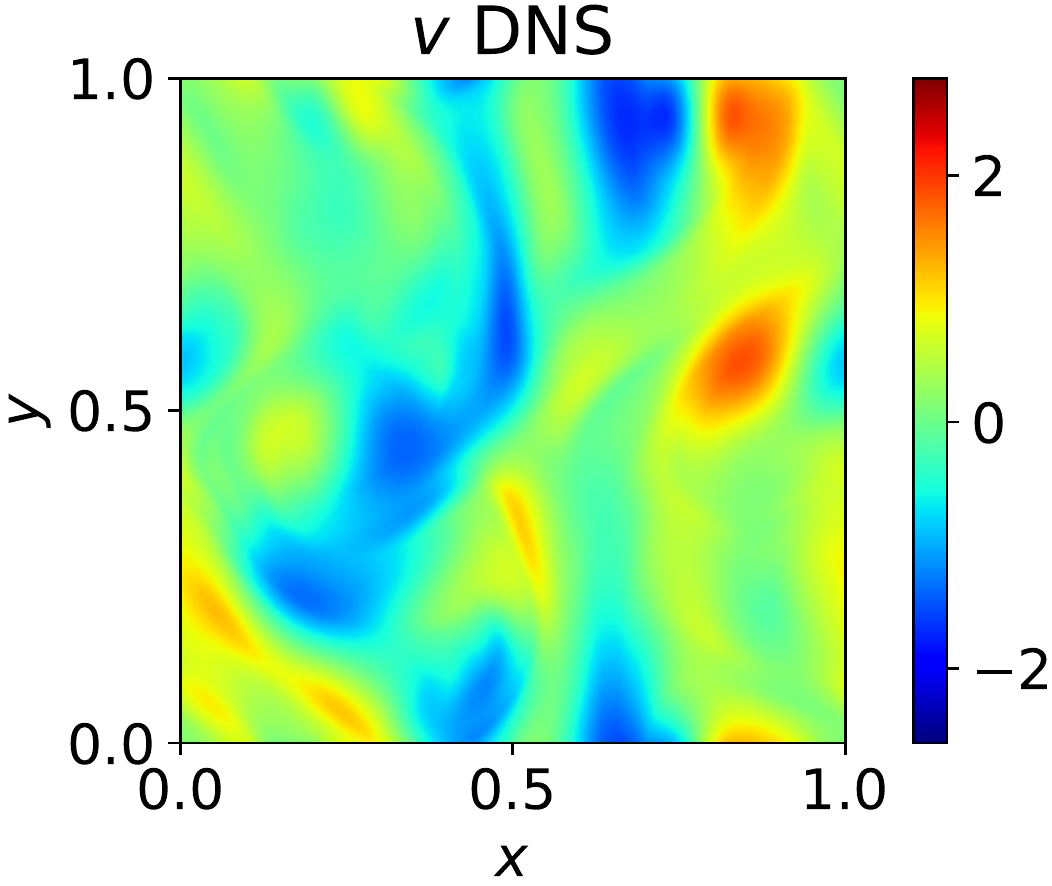}
\end{subfigure}
\begin{subfigure}[t]{0.33\textwidth}
\includegraphics[width=\textwidth,height=\textheight,keepaspectratio]{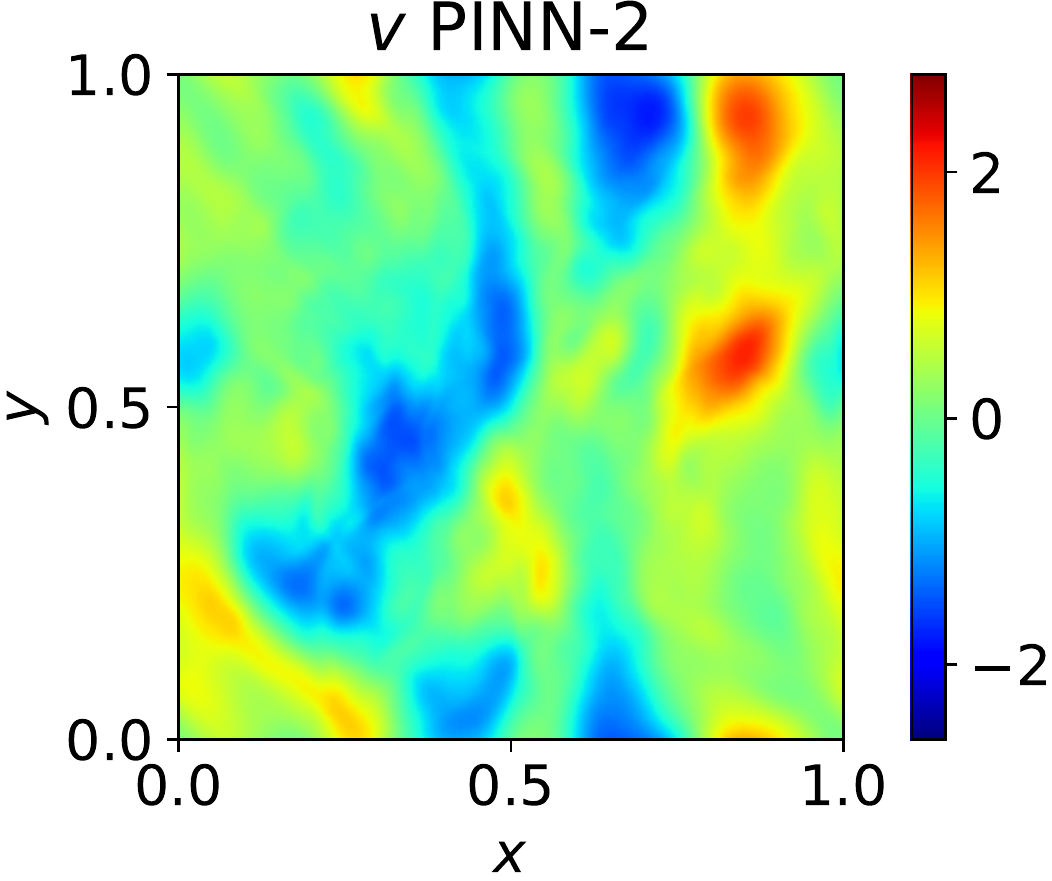}
\end{subfigure}
\begin{subfigure}[t]{0.33\textwidth}
\includegraphics[width=\textwidth,height=\textheight,keepaspectratio]{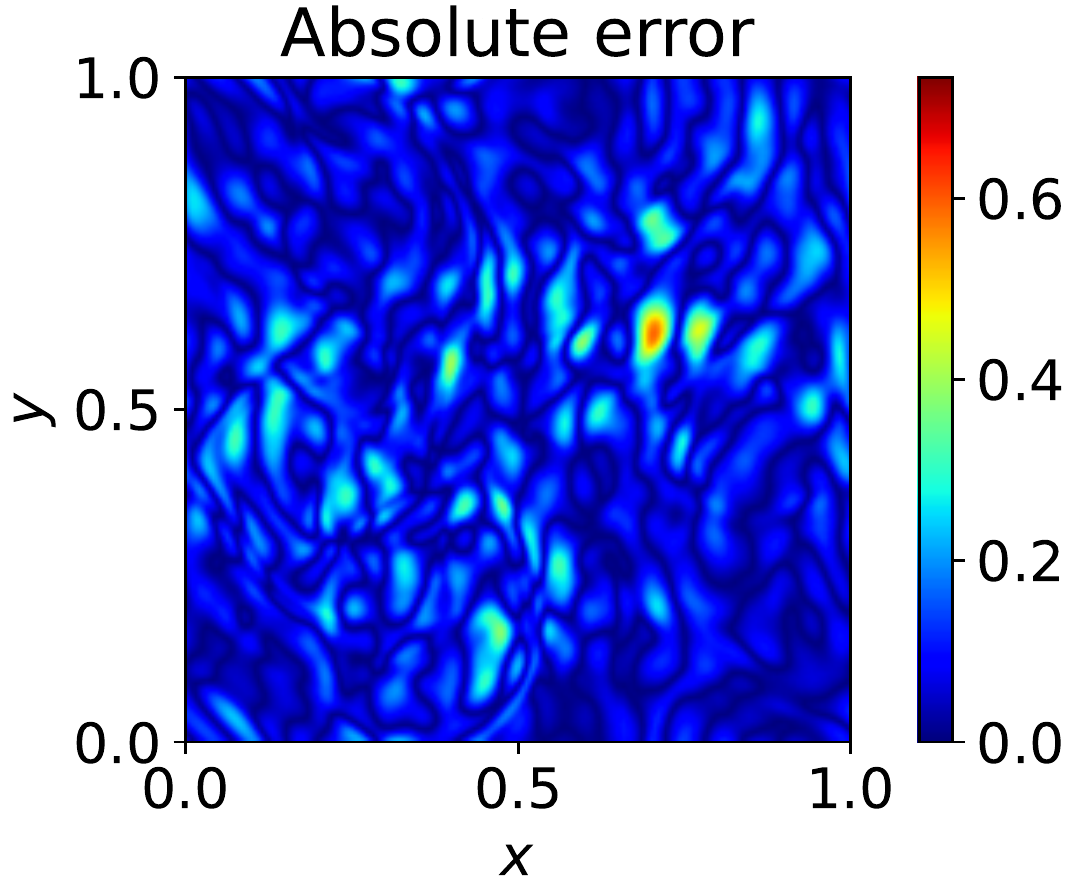}
\end{subfigure}
\caption{\label{fig:wide} Contour plots of the velocity components in comparison with DNS at time $t=0.45$.}
\label{fig:contours_t1}
\end{figure*}
\begin{figure*}
\begin{subfigure}[t]{0.33\textwidth}
\includegraphics[width=\textwidth,height=\textheight,keepaspectratio]{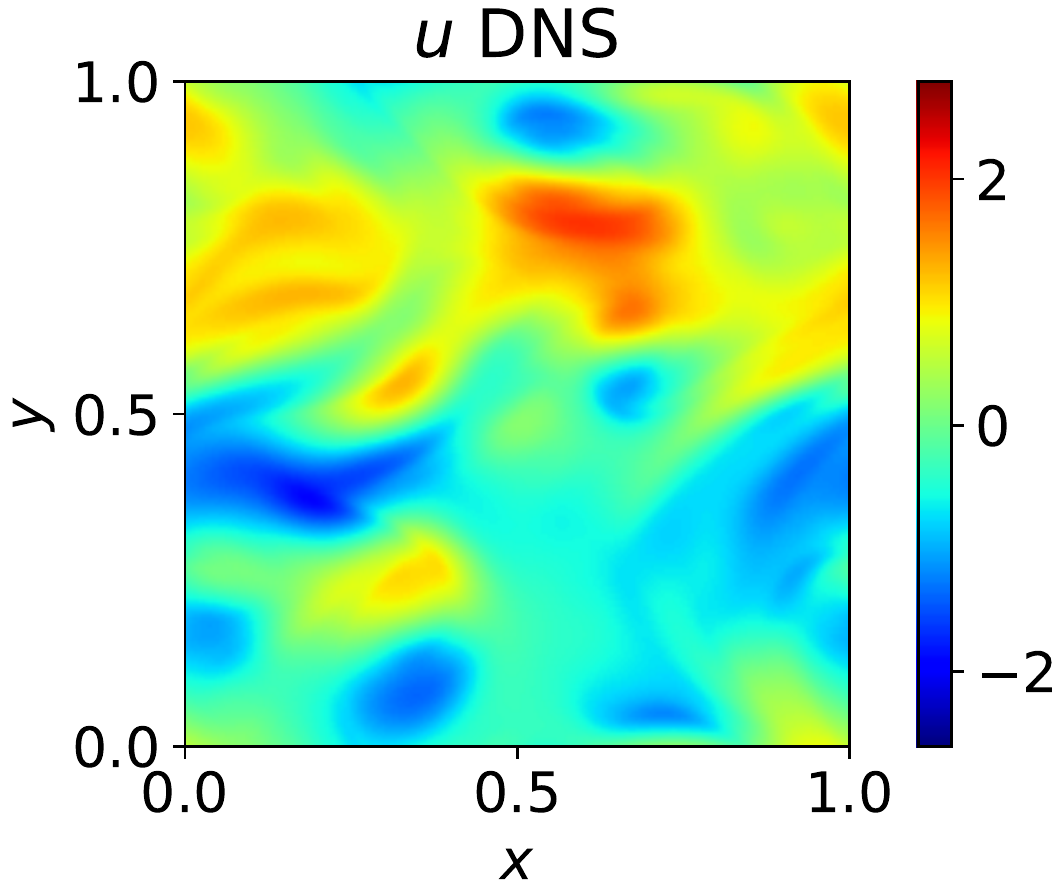}
\end{subfigure}
\begin{subfigure}[t]{0.33\textwidth}
\includegraphics[width=\textwidth,height=\textheight,keepaspectratio]{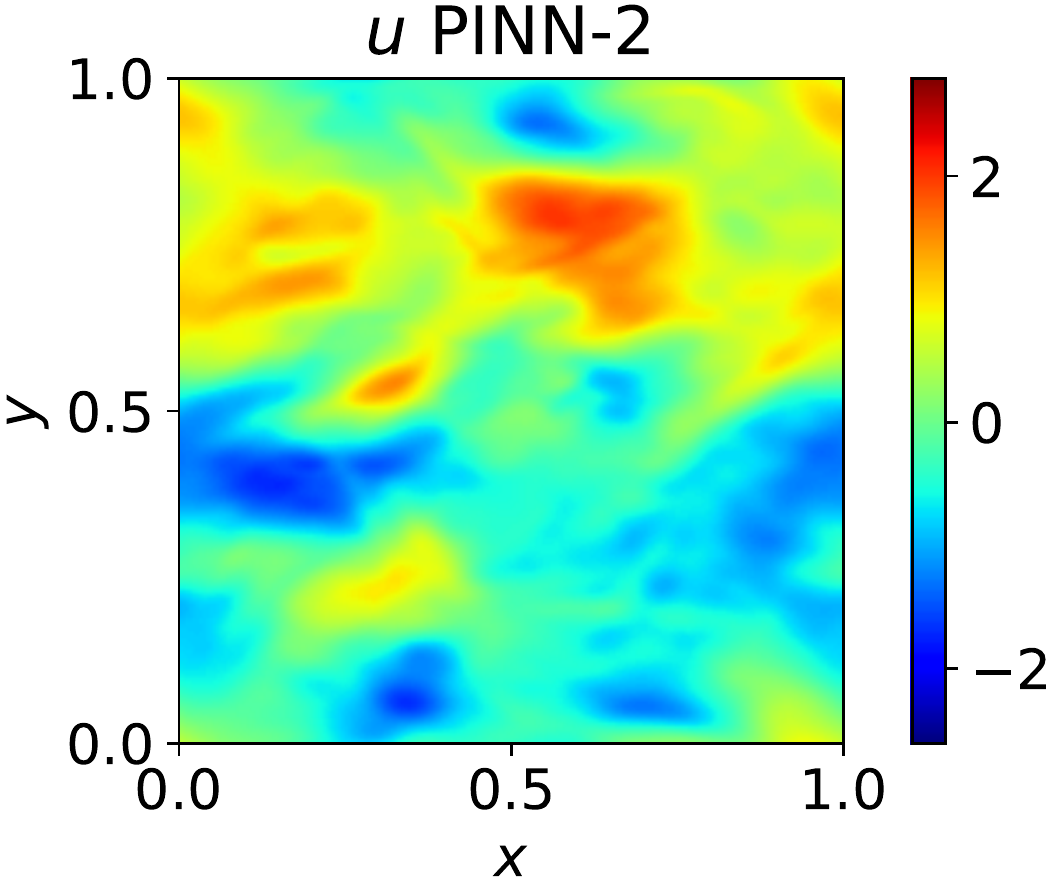}
\end{subfigure}
\begin{subfigure}[t]{0.33\textwidth}
\includegraphics[width=\textwidth,height=\textheight,keepaspectratio]{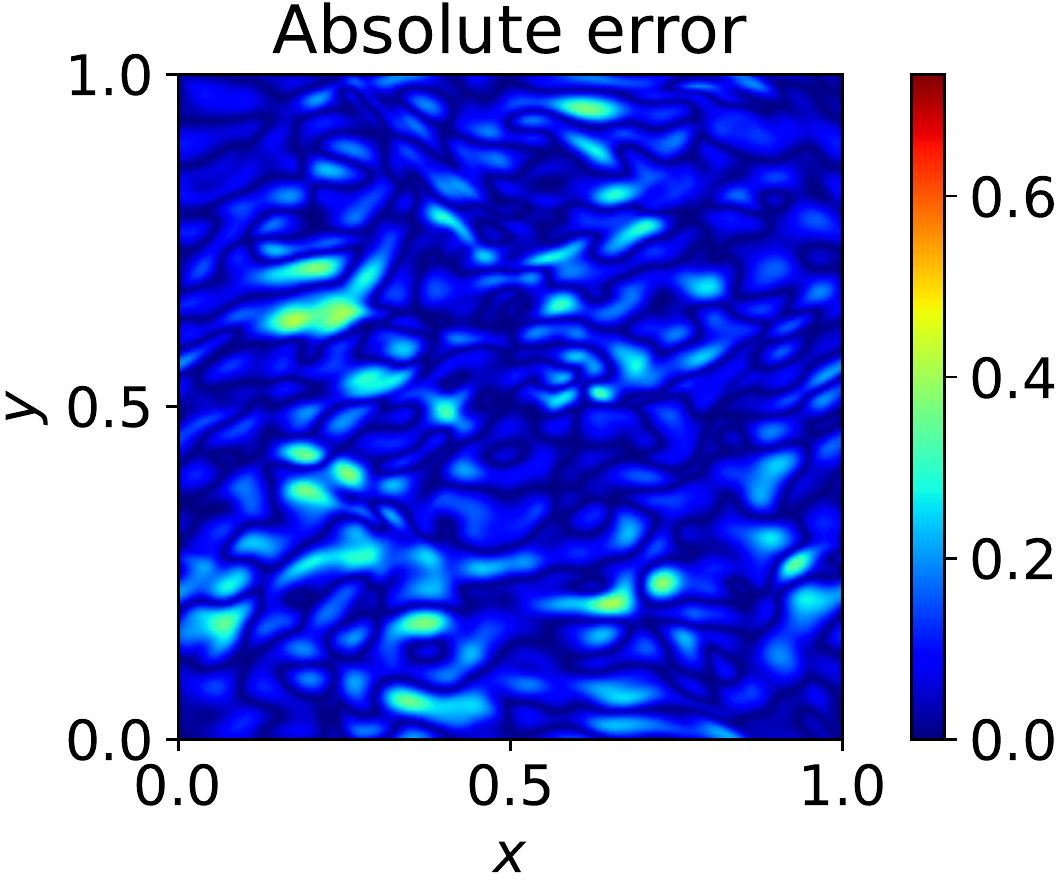}
\end{subfigure}
\begin{subfigure}[t]{0.33\textwidth}
\includegraphics[width=\textwidth,height=\textheight,keepaspectratio]{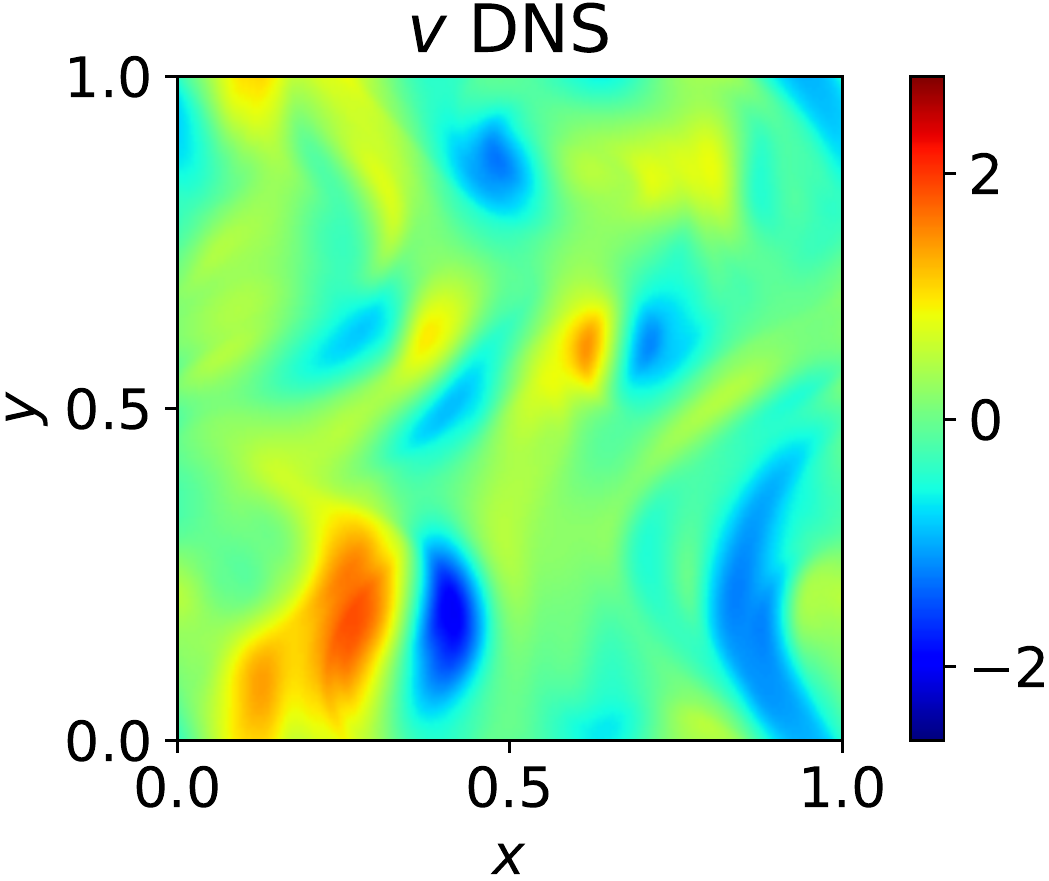}
\end{subfigure}
\begin{subfigure}[t]{0.33\textwidth}
\includegraphics[width=\textwidth,height=\textheight,keepaspectratio]{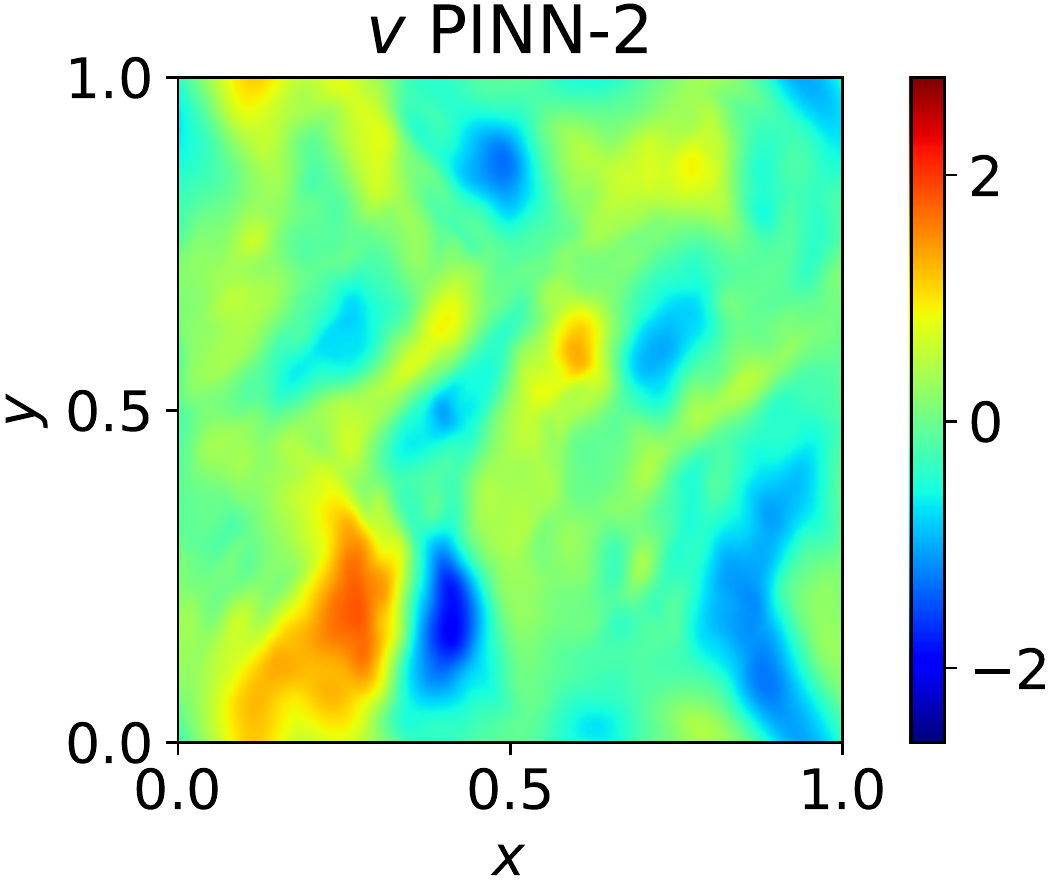}
\end{subfigure}
\begin{subfigure}[t]{0.33\textwidth}
\includegraphics[width=\textwidth,height=\textheight,keepaspectratio]{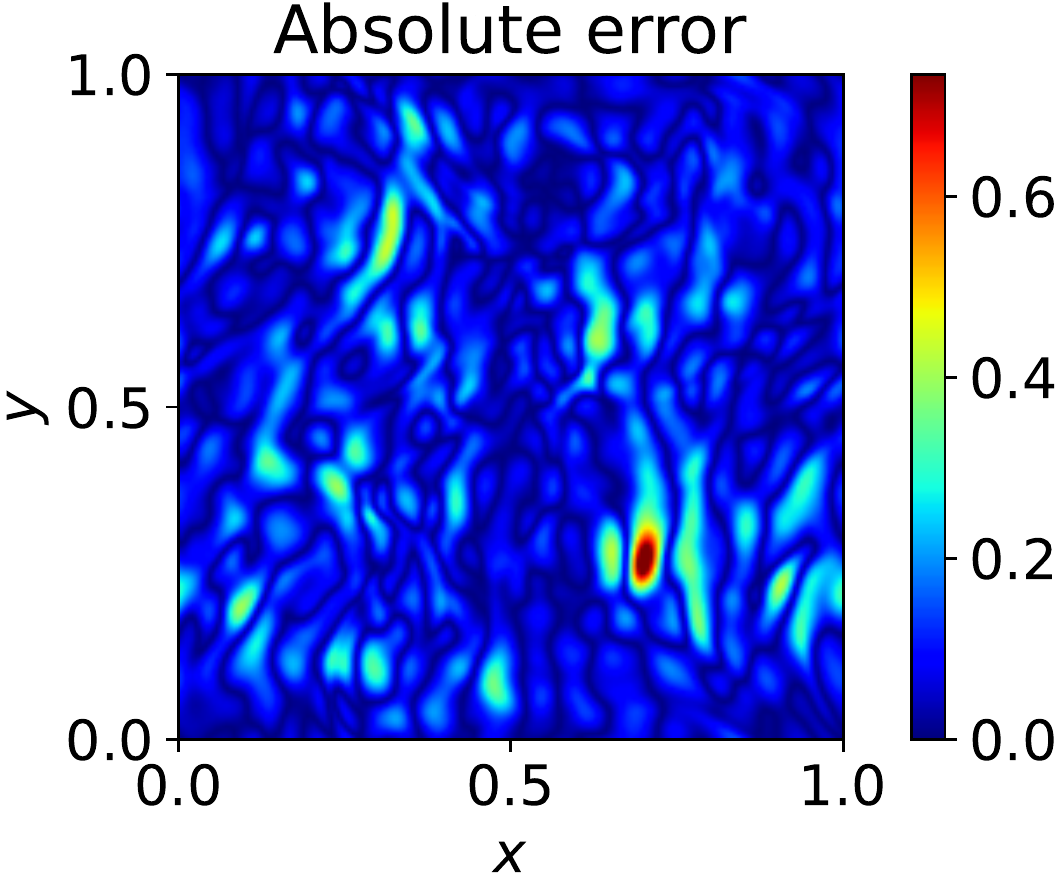}
\end{subfigure}
\caption{\label{fig:wide} Contour plots of the velocity components in comparison with DNS at time $t=0.98$.}
\label{fig:contours_t2}
\end{figure*}

\subsection{Remarks on PINNs for turbulence simulations}

The PINN-2 model compares well with the DNS results both at testing and training times. 
From the spectra we see that the model works best at large scales and to study this further we look at the temporal fluctuations of some of the large scale modes. 
We denote $\hat{\psi} ({\bf k}, t)$ as the two-dimensional Fourier transform of the streamfunction $\psi (x, y, t)$. 
In Fig. \ref{fig:large_modes} we show the time series of the real and imaginary part of the modes $\hat{\psi}_{k_x,k_y}$ for a few values of wavenumbers $k_x, k_y$ for both the PINN-2 model and the DNS. 
The symbol $\hat{\psi}^r$ denotes the real part of the mode $\hat{\psi}$ while the symbol $\hat{\psi}^i$ denotes the imaginary part of the mode $\hat{\psi}$.
As seen from the plots, the fluctuations of the large scale modes $\hat{\psi}_{11}, \hat{\psi}_{12}, \hat{\psi}_{21}, {\markup{\hat{\psi}_{22}}}$, are captured very well by the PINN-2 model.  
For smaller scale modes $\hat{\psi}_{33}, \hat{\psi}_{44}$ we see that the time series starts to deviate from the DNS results. 
Since the largest modes are accurately captured, we can infer that the PINN-2 model learns to capture the long range correlations more easily than the short range correlations. 
While for the small scales, the statistical behaviour of these modes like the typical fluctuation amplitudes, are captured by the PINN-2 model as seen in Figs. \ref{fig:large_modes}.  
\begin{figure*}
\begin{subfigure}[t]{0.33\textwidth}
\includegraphics[width=\textwidth,height=\textheight,keepaspectratio]{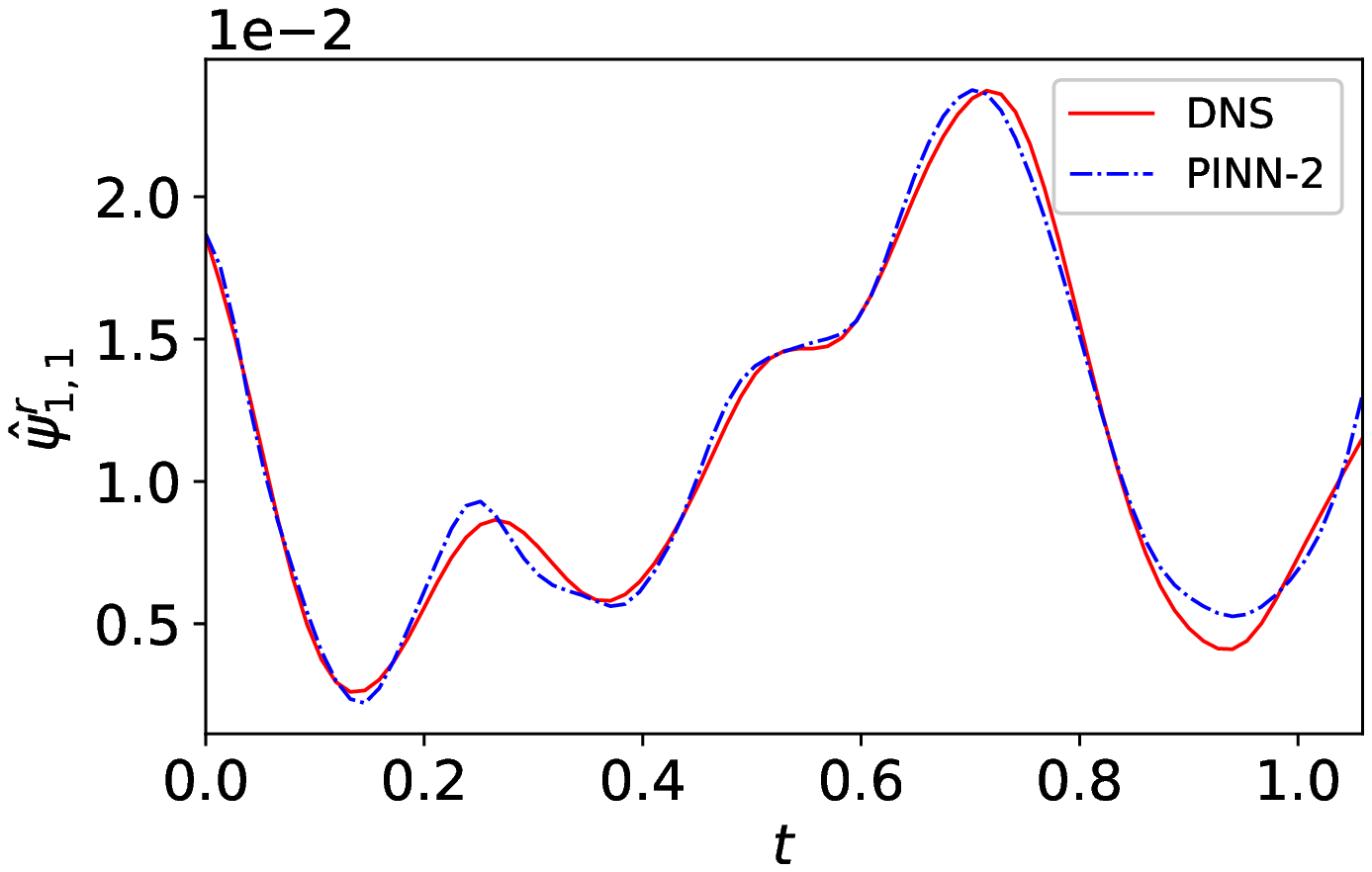}
\end{subfigure}
\begin{subfigure}[t]{0.33\textwidth}
\includegraphics[width=\textwidth,height=\textheight,keepaspectratio]{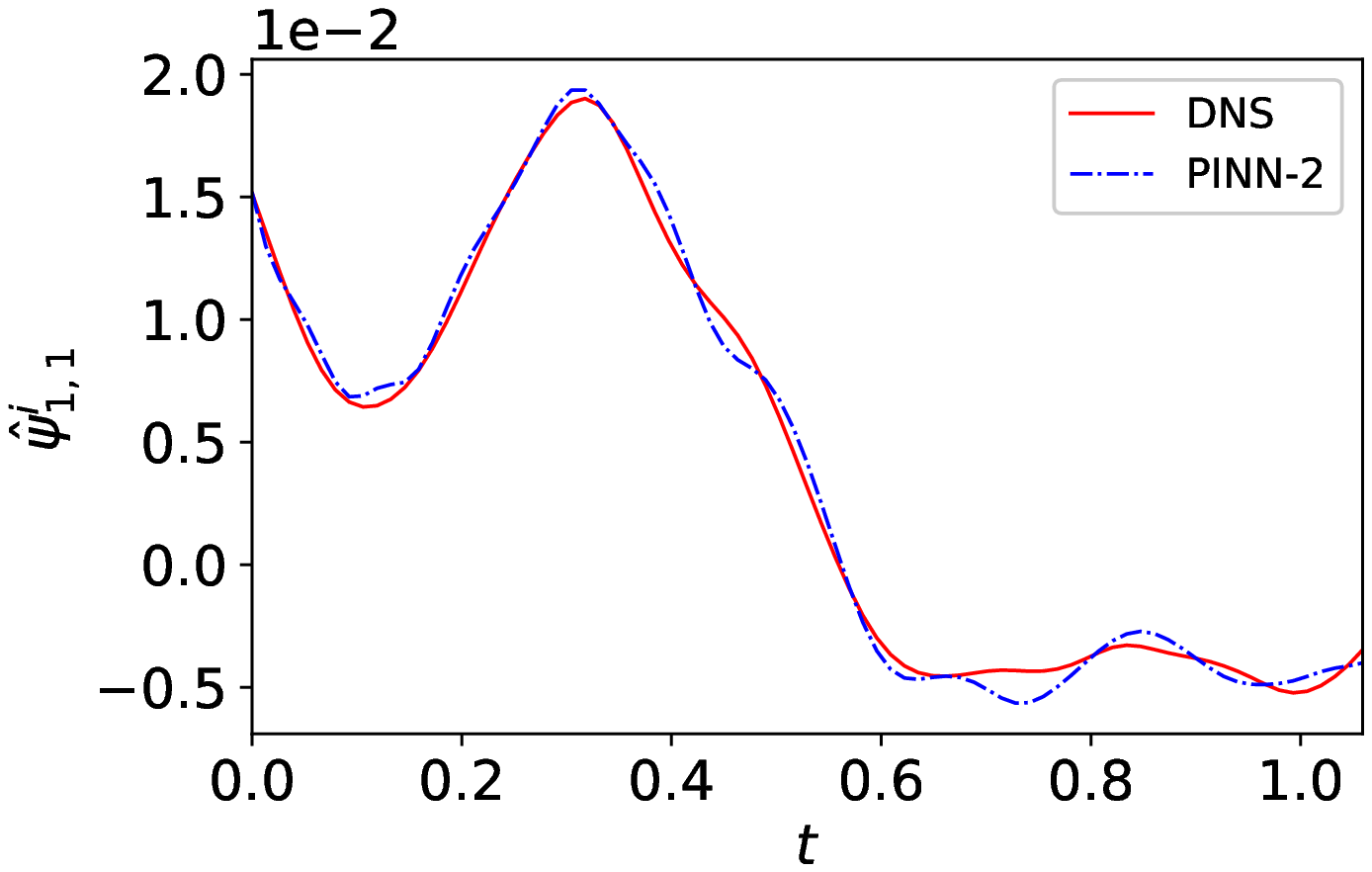}
\end{subfigure}
\begin{subfigure}[t]{0.33\textwidth}
\includegraphics[width=\textwidth,height=\textheight,keepaspectratio]{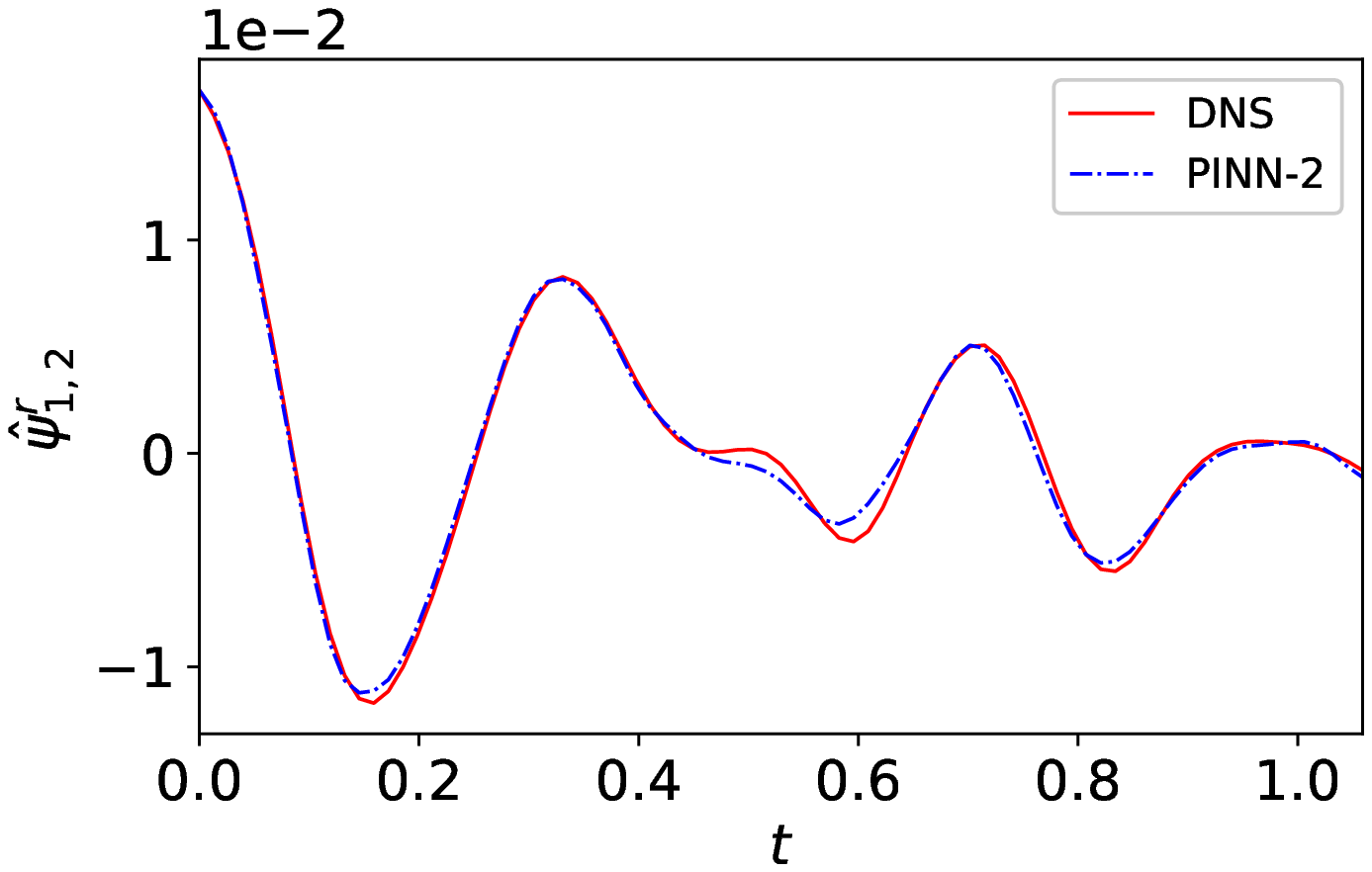}
\end{subfigure}
\begin{subfigure}[t]{0.33\textwidth}
\includegraphics[width=\textwidth,height=\textheight,keepaspectratio]{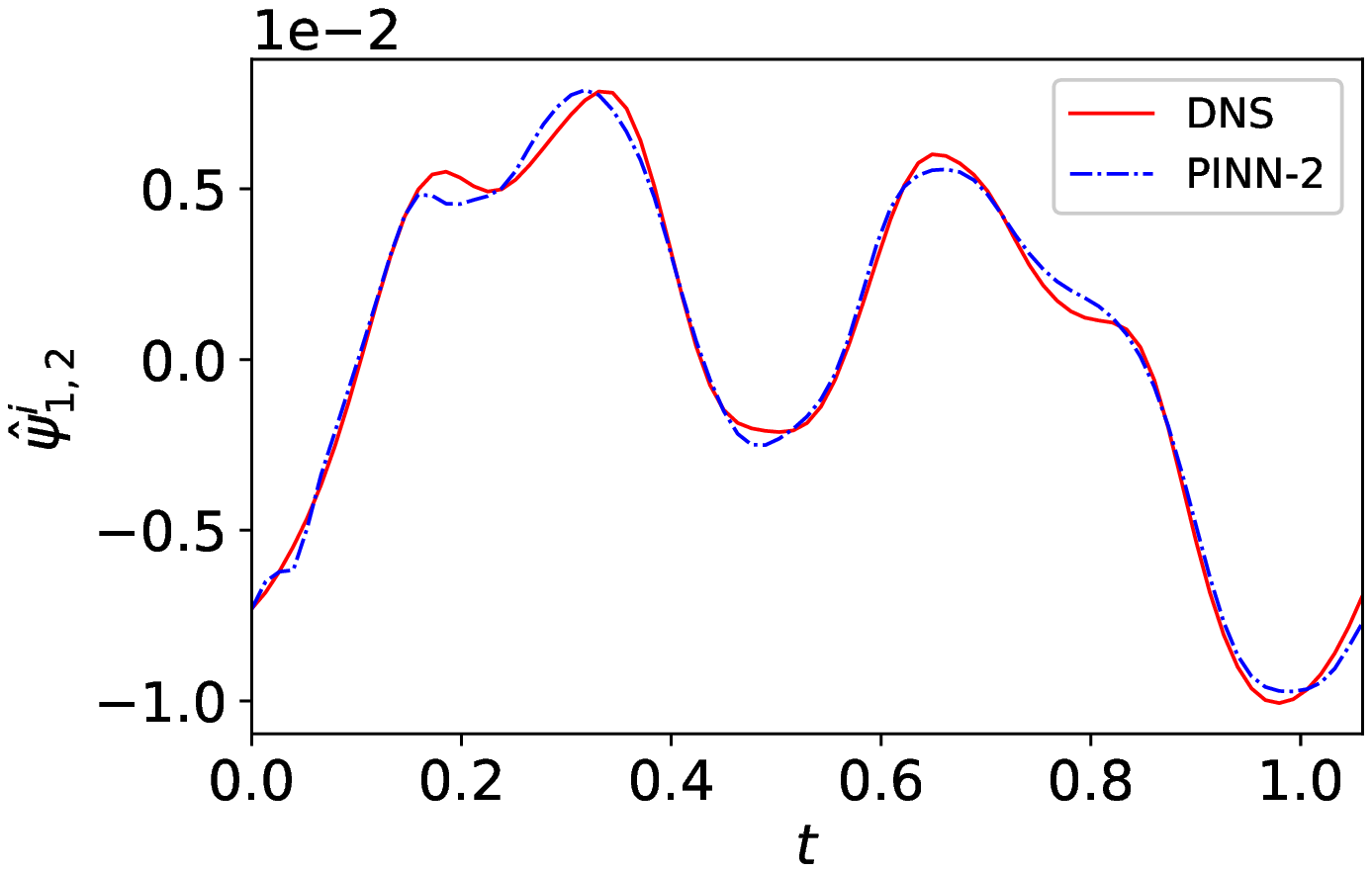}
\end{subfigure}
\begin{subfigure}[t]{0.33\textwidth}
\includegraphics[width=\textwidth,height=\textheight,keepaspectratio]{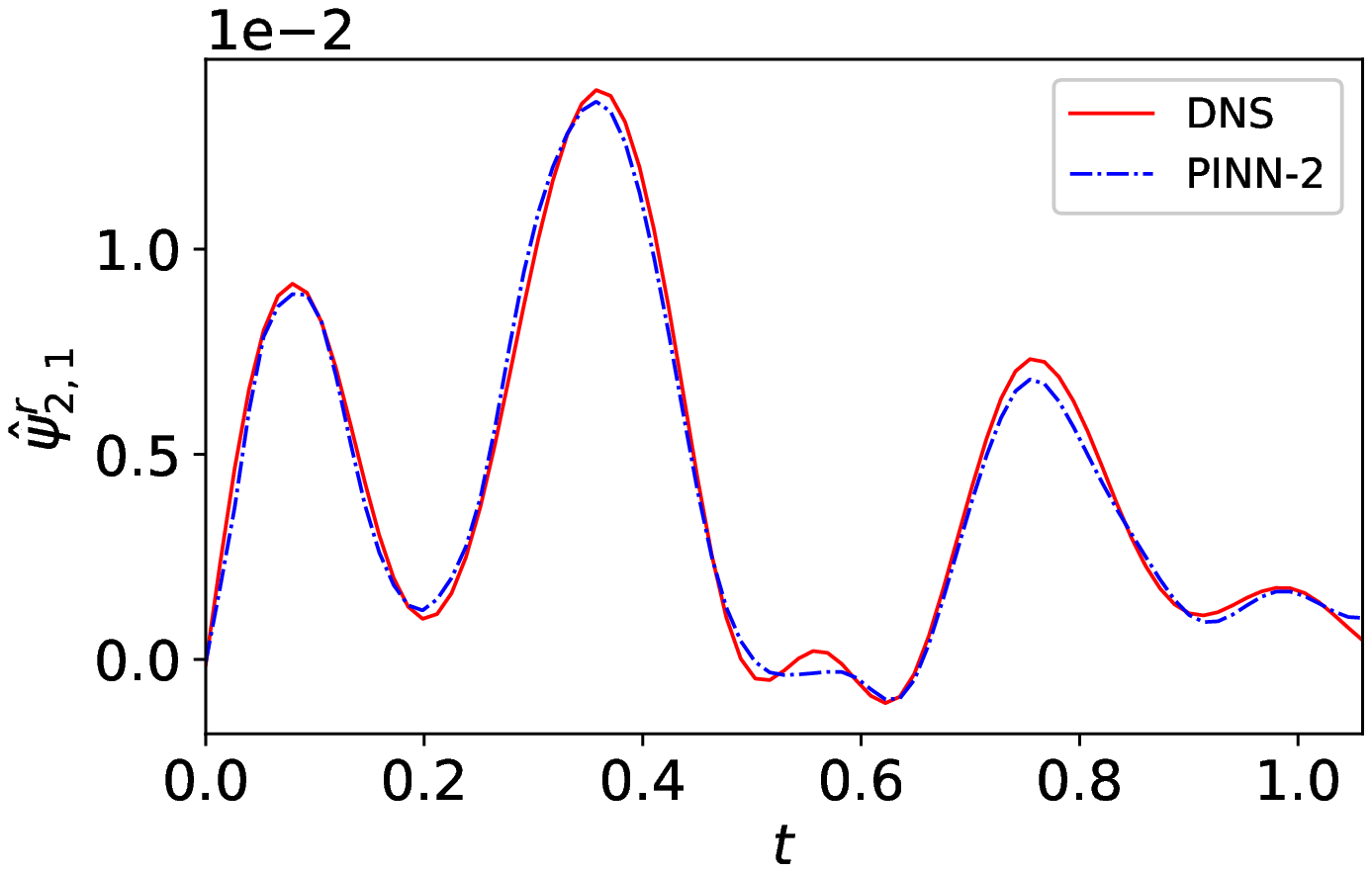}
\end{subfigure}
\begin{subfigure}[t]{0.33\textwidth}
\includegraphics[width=\textwidth,height=\textheight,keepaspectratio]{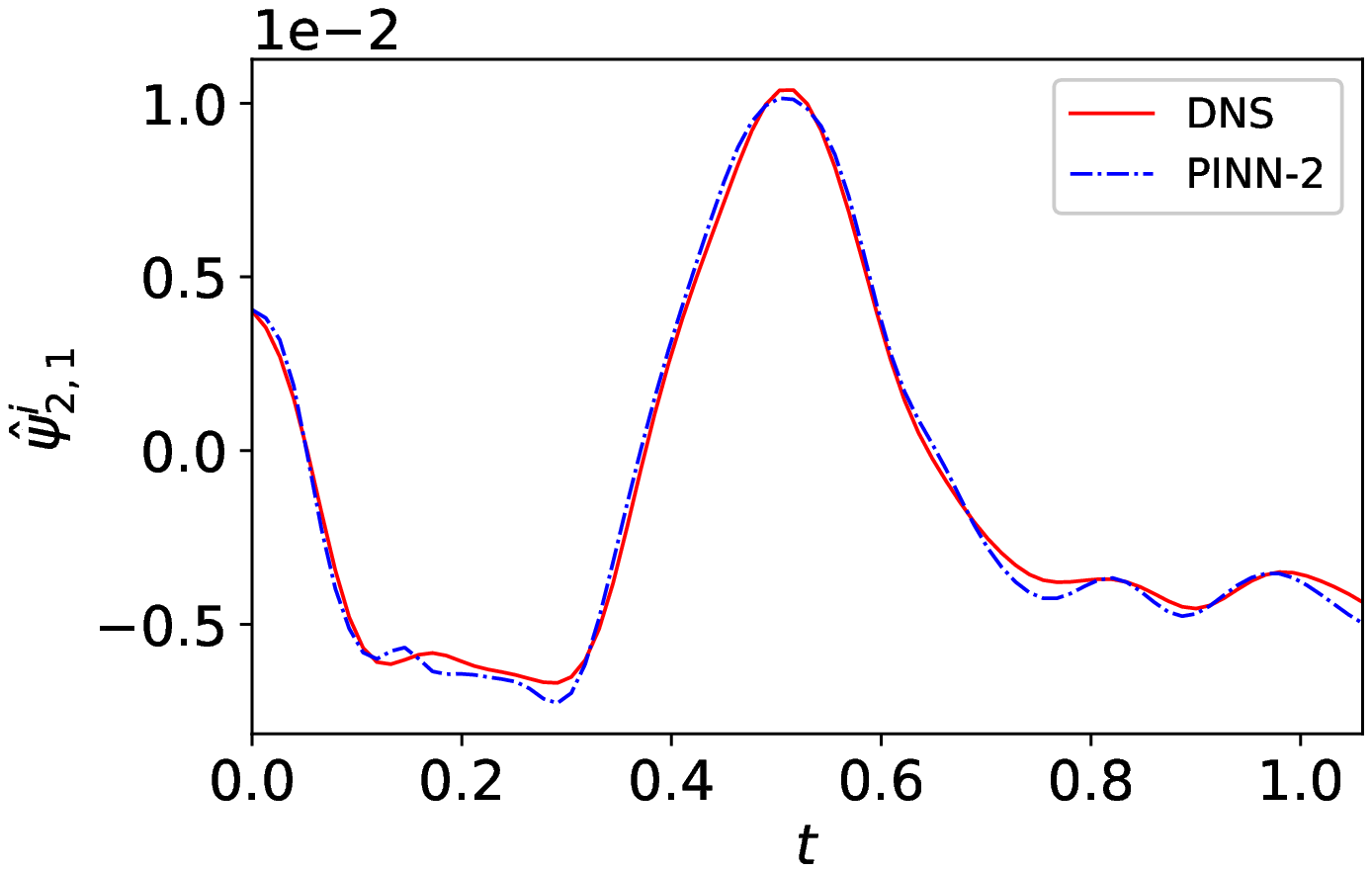}
\end{subfigure}
\begin{subfigure}[t]{0.33\textwidth}
\includegraphics[width=\textwidth,height=\textheight,keepaspectratio]{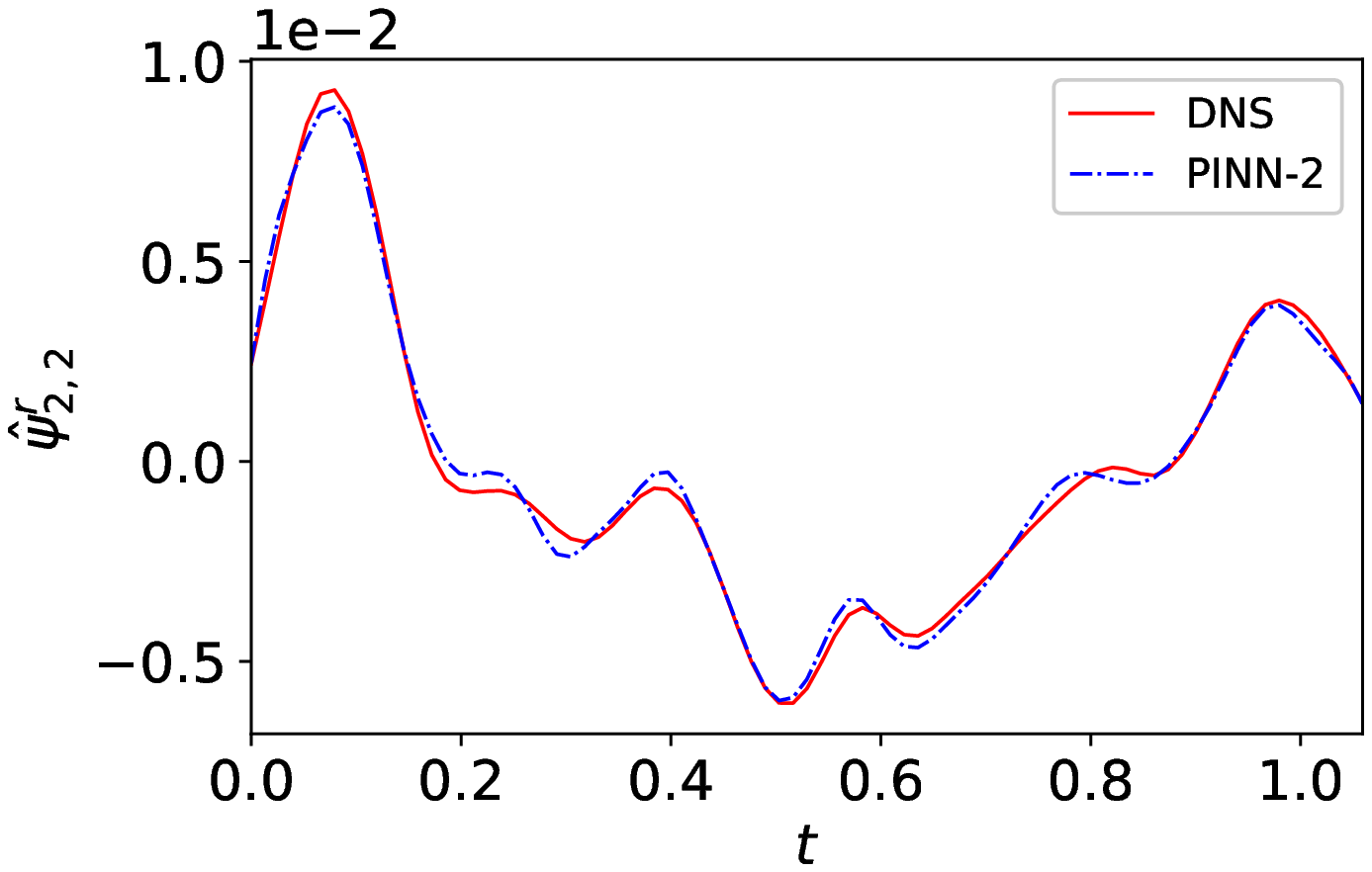}
\end{subfigure}
\begin{subfigure}[t]{0.33\textwidth}
\includegraphics[width=\textwidth,height=\textheight,keepaspectratio]{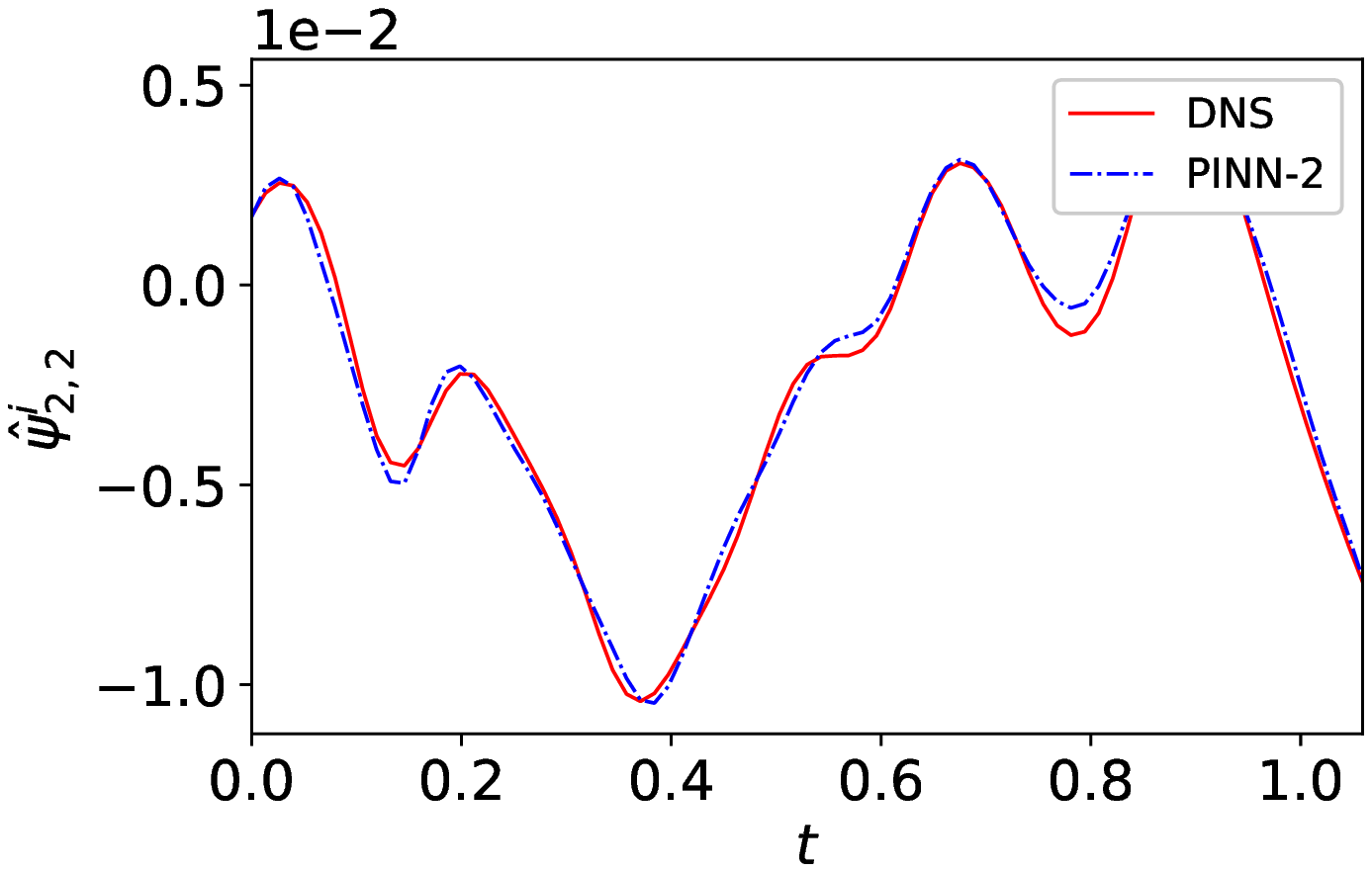}
\end{subfigure}
\begin{subfigure}[t]{0.33\textwidth}
\includegraphics[width=\textwidth,height=\textheight,keepaspectratio]{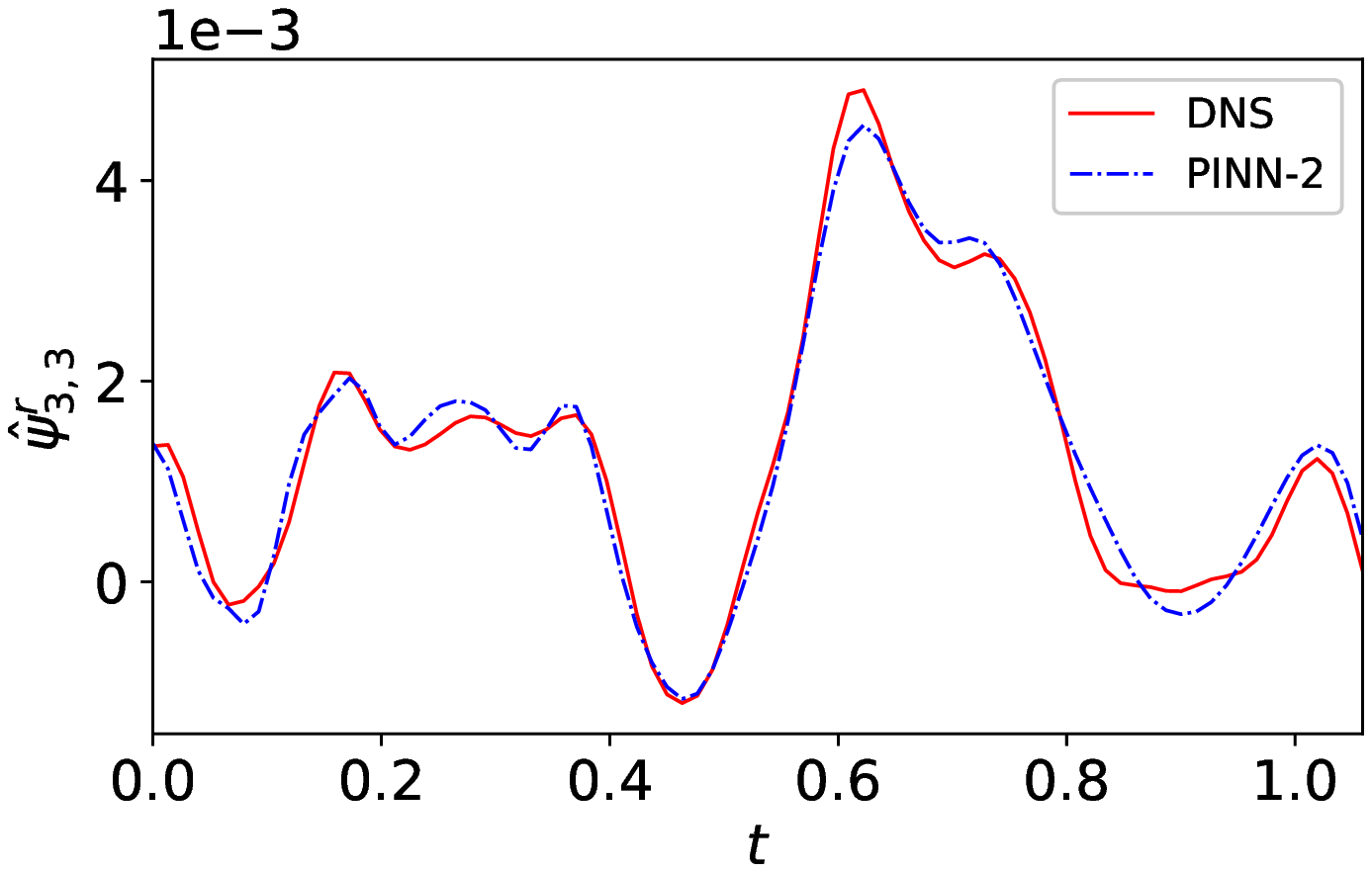}
\end{subfigure}
\begin{subfigure}[t]{0.33\textwidth}
\includegraphics[width=\textwidth,height=\textheight,keepaspectratio]{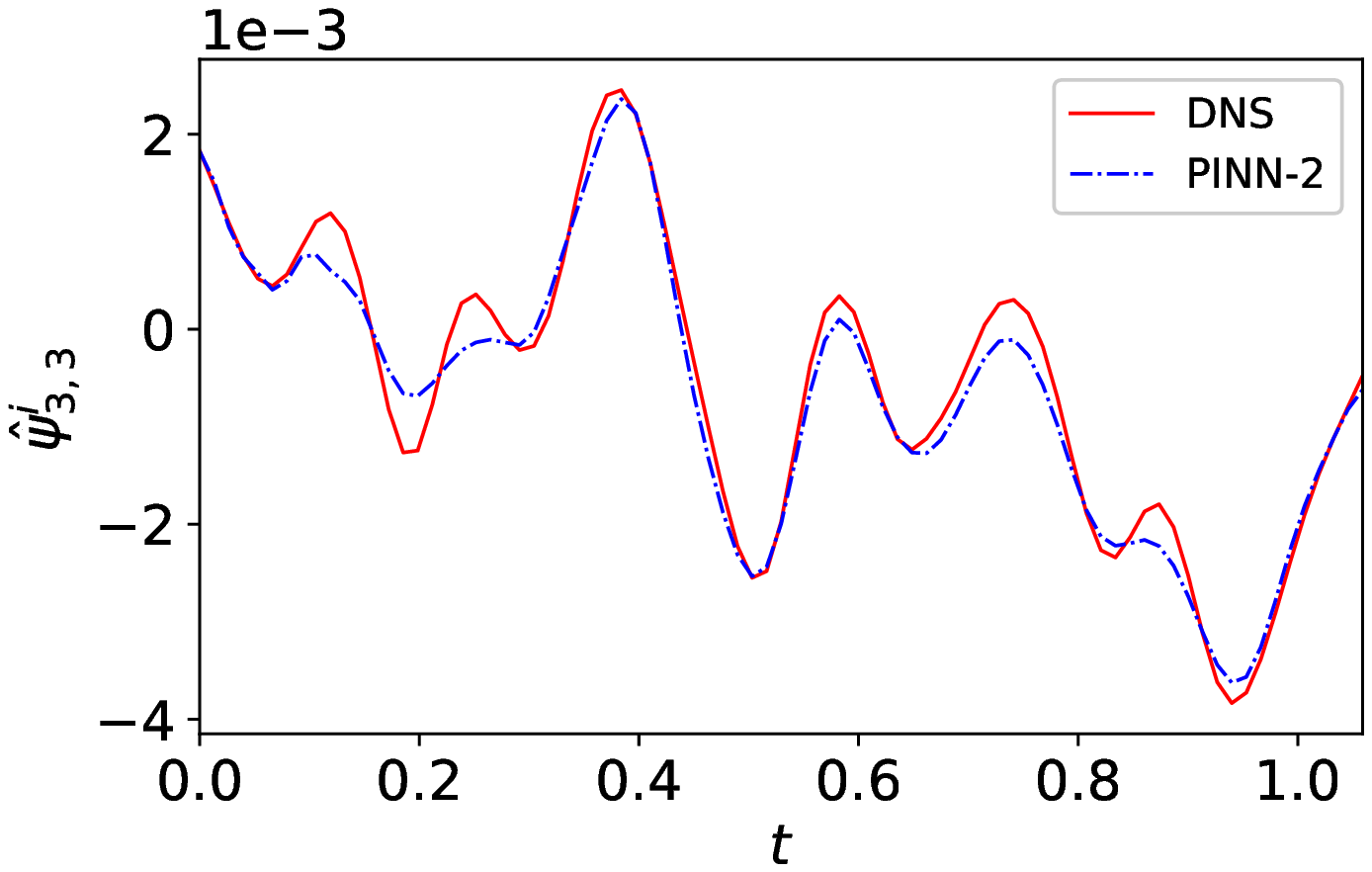}
\end{subfigure}
\begin{subfigure}[t]{0.33\textwidth}
\includegraphics[width=\textwidth,height=\textheight,keepaspectratio]{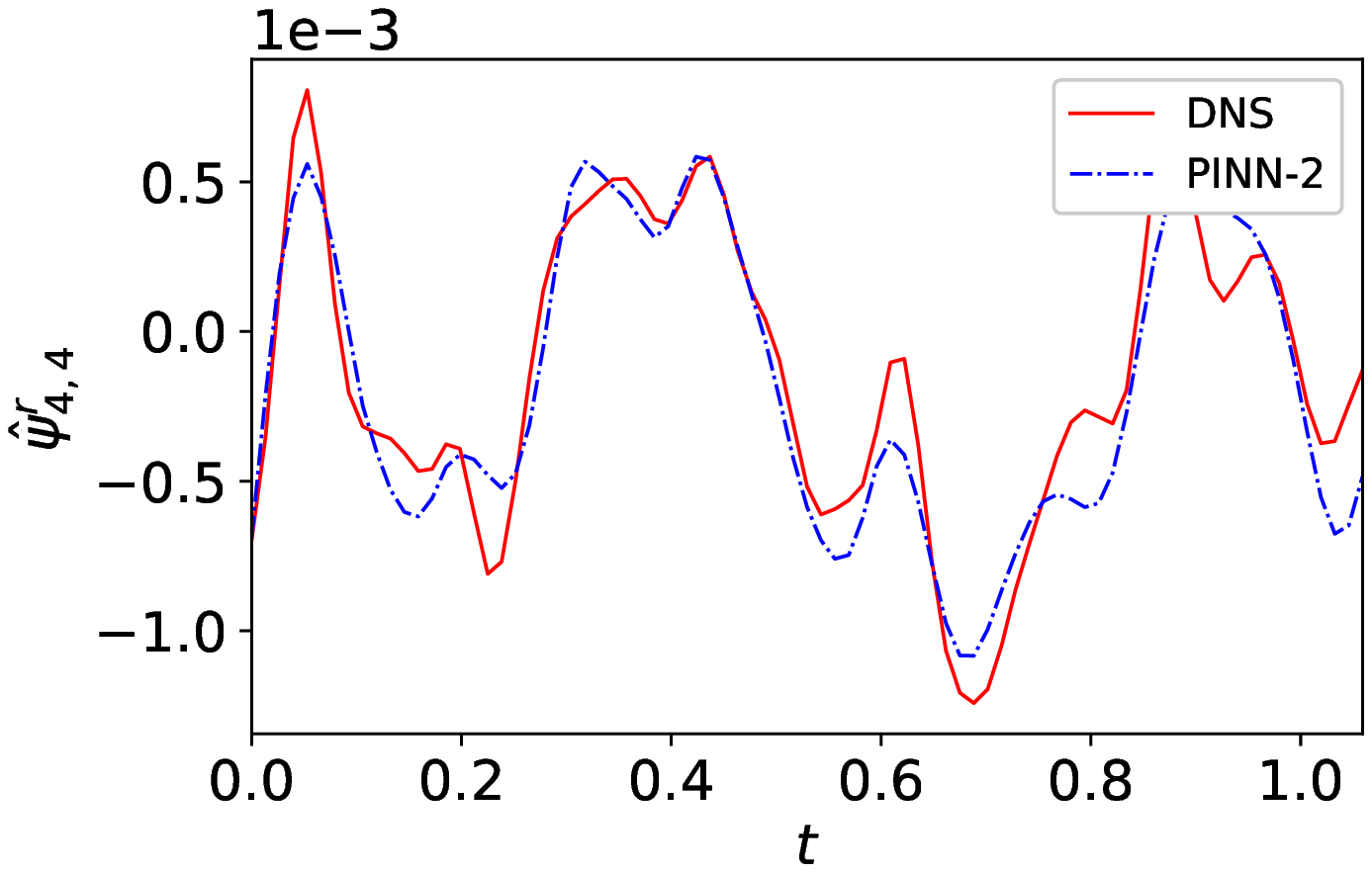}
\end{subfigure}
\begin{subfigure}[t]{0.33\textwidth}
\includegraphics[width=\textwidth,height=\textheight,keepaspectratio]{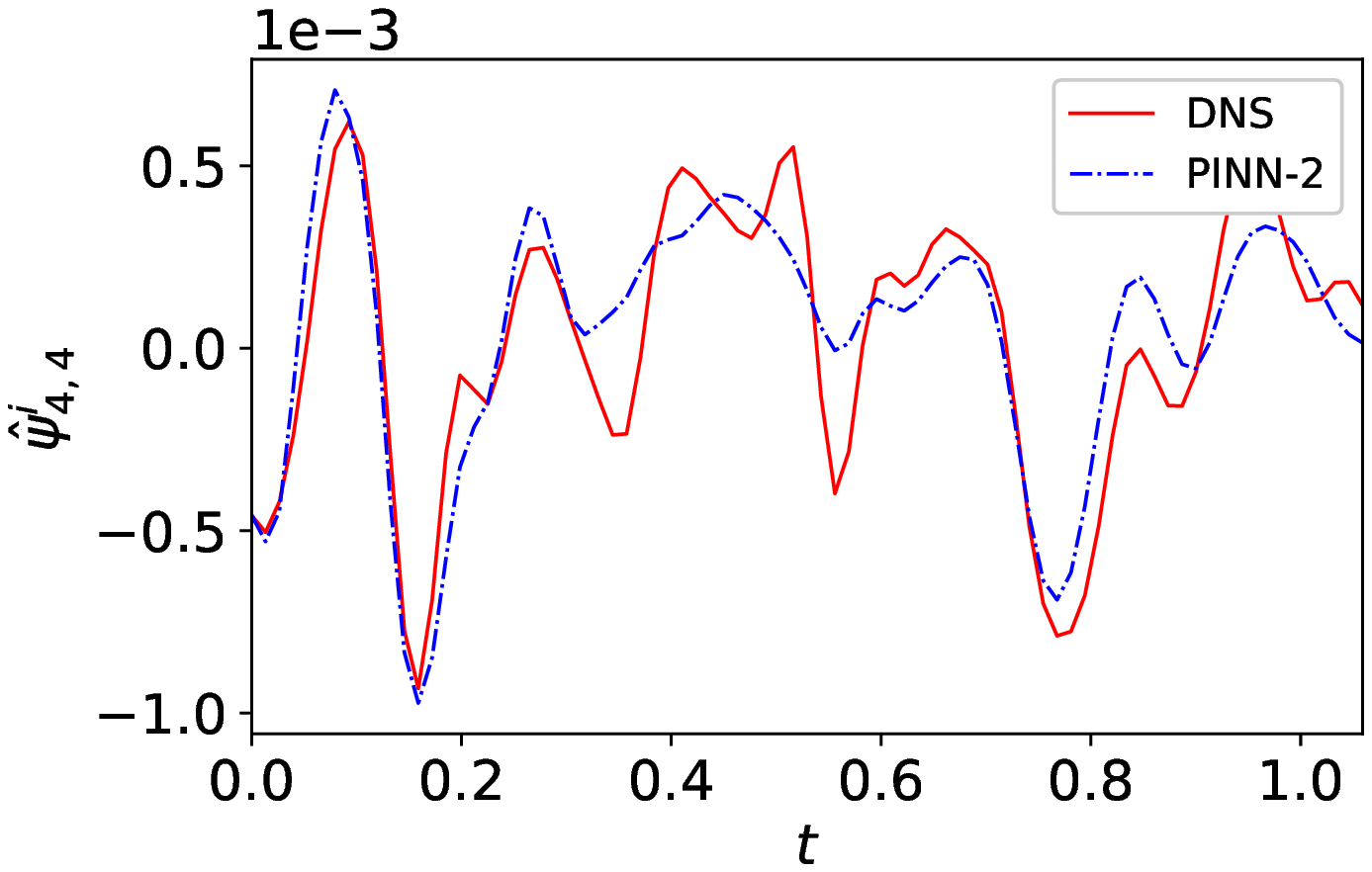}
\end{subfigure}
\caption{\label{fig:wide} {\markup{Time series}} of the real and complex part of the amplitudes of different Fourier modes $\hat{\psi}$ for both the PINN-2 model and the DNS.}
\label{fig:large_modes}
\end{figure*}

We thus find that the PINN-2 model captures \markup{very well the large scale features of the two-dimensional turbulent flow even though the temporal behaviour of the small scale modes are found to deviate from the DNS results. This happens since the small scale modes only act as an effective viscosity on the large scale modes}. In this particular study we have used DNS results to train our PINN model, but we can in principle also use experimental data from Particle Image Velocimetry (PIV) measurements. Such usage of PINNs has already been shown to be successful in the works of \cite{hfm20,cai2021}, and it reinforces our belief that the results from the PINN-2 model can be used to study even turbulent flows with larger Reynolds numbers than currently used in this study. 

\newpage
\section{Conclusion}

We use two PINN models to predict {\markup{two-dimensional turbulence}} in this study. The first approach (PINN-1) involves a standard PINN model \citep{raissi19} along with training of solution data from the interior of the domain.
The second approach (PINN-2) is similar but involves two neural networks (low and high wavenumbers) where the initial and data losses are minimized separately for these networks. We perform a systematic hyperparameter search and we obtain an optimal set for which the total loss value is minimal. We observe that PINN-2 outperforms PINN-1 with a lower RMSE value and higher $R^2$ score. Furthermore, we penalize the equation loss to get physically acceptable solutions, which also allows us to capture the energy spectra more accurately. The amount of training data that is used to train these models is close to $0.1\%$ of the whole data set available at training time instances. 
The errors and comparisons are done at testing time instances {\markup{which is three times the number of training time instances}}. 

We observe that PINN-2 model predicts the velocities to reasonable accuracy as compared to the DNS results at the testing times. It provides a solution that captures the fluctuations of different quantities and their time averaged behaviour based upon the constraints from DNS data, the governing equations, boundary conditions and initial conditions. Since we use only $0.1 \%$ of the data from DNS for training the PINN based model, the results open up the possibilities of using such PINN models for studying turbulence at very fine time intervals, where standard DNS simulations face storage issues. The PINN-2 model studied here shows a very good ability to capture the fluctuations of large scale modes. Such models can possibly help in studying long time statistics of systems that show phenomena such as flow reversals and $1/f$-noise \citep{herault20151,dallas2020transitions}. 

{\markup{PINN-2 model utilizes a Fourier filter based domain decomposition to filter out and train the low and high wavenumber components of the quantity of interest}}. For future work, following the design of PINN-2 model, {\markup{one could}} expand it further by adding additional neural networks to the existing architecture. {\markup{Furthermore, one can in principle use such a network for multiscale physical problems beyond fluid dynamics}}.  Also, {\markup{one could}} introduce adaptive weights for the different loss functions which can in principle improve the predictive accuracy \citep{wang20}. Since PINN-2 model is able to capture energy spectral density across many decades, while falling into the length scales where transition between inertial and viscous range occurs, {\markup{one can}} further explore more sophisticated models that can capture the whole viscous range accurately. Such models can also help in studying systems where DNS simulations are very expensive or difficult to be carried out. 


\section*{Acknowledgements}
The authors thank the three referees whose insightful comments and suggestions helped improve this manuscript. The authors thank the computing resources and support provided by the CI/PDP Deep Learning Platform Team at Robert Bosch GmbH, Renningen, Germany and the PARAM Shakti supercomputing facility of IIT Kharagpur established under National Supercomputing Mission (NSM), Government of India and supported by Centre for Development of Advanced Computing (CDAC), Pune. K. Seshasayanan acknowledges support from NSM Grant no. DST/NSM/R\&D\_HPC\_Applications/2021/03.11, from the Institute Scheme from Innovative Research and Development (ISIRD), IIT Kharagpur, Grant No. IIT/SRIC/ISIRD/2021-2022/08 and the Start-up Research Grant No. SRG/2021/001229 from Science \& Engineering Research Board (SERB), India. 

\section*{Data Availability Statement}

The data that support the findings of this study are available from the corresponding author upon reasonable request.

\bibliography{Article_file}
\bibliographystyle{abbrvnat2}

\end{document}